\documentclass[12pt,a4paper]{article}

\usepackage[T1]{fontenc}
\usepackage[utf8]{inputenc}
\usepackage{geometry}
\usepackage{setspace}
\usepackage{amsmath}
\usepackage{amssymb}
\usepackage{amsfonts}
\usepackage{bm}
\usepackage{graphicx}
\usepackage{booktabs}
\usepackage[dvipsnames]{xcolor}
\usepackage[numbers,sort&compress]{natbib}
\usepackage[colorlinks=true,citecolor=blue,linkcolor=blue,urlcolor=blue]{hyperref}
\usepackage[colorinlistoftodos,prependcaption,textsize=footnotesize]{todonotes}
\usepackage{authblk}

\newcommand{\bvec}[1]{\bm{#1}}
\newcommand{\sigT}{\sigma_{\mathrm{T}}}
\newcommand{\rhog}{\bar{\rho}_{\gamma}}

\newcommand{\Hubble}{\mathcal{H}}

\begin{document}
%%%%%%%%%%%%%%%%%%%%%%%%%%%%%%%%%%%%%%%%%%%%%%%%%%%%%%%%%%%%%%%%%%%%%%%%%%%%%

\title{\Large\bfseries Kinetic Theory of Cosmological Magnetogenesis at Second Order: A New Density-Gradient Source and Comparison with the Harrison Mechanism}

\author[1,2]{Bob Osano}
\affil[1]{\small Cosmology and Gravity Group, Department of Mathematics and
Applied Mathematics, University of Cape Town, Rondebosch 7701, South Africa}
\affil[2]{\small Centre for Higher Education Development,
University of Cape Town, Rondebosch 7701, South Africa}
\date{Correspondence: \texttt{bob.osano@uct.ac.za}\\[4pt]\today}

\maketitle
\thispagestyle{empty}

\begin{abstract}
\noindent
We derive and compare three mechanisms of cosmological magnetogenesis: the
Thomson-scattering velocity-difference mechanism of Takahashi et al.\ (2005),
a new density-gradient source identified here for the first time, and the
Harrison bulk-flow mechanism of Cembranos et al.\ (2020).  Starting from the
coupled Maxwell--Boltzmann equations, the complete kinetic theory chain is
derived in a single document---from the BBGKY hierarchy and Thomson collision
term, through the generalised Ohm's law, to the second-order magnetic induction
equation.  The Ohm's law correction terms are each bounded by
$m_e/m_p\approx5.4\times10^{-4}$, confirming the standard single-fluid
approximation to better than $0.1\%$.  At second order in cosmological
perturbations, products of first-order scalar source vorticity; we identify
a coupling between the photon density contrast
$\delta_\gamma\equiv\delta\rho^{(1)}_\gamma/\bar\rho_\gamma$ and the
electron--photon velocity difference $(u_e-u_\gamma)^{(1)}$ that was implicitly
present in previous treatments but never isolated.  Numerical evaluation with
CAMB~v1.6.6 at $z=1100$ shows that this term contributes at
${\approx}0.97\times B_{\rm Tak}$, giving a scattering-mechanism total
${\approx}1.4\times$ the Takahashi result.  The Harrison mechanism at the
Planck bulk-flow limit ($\beta<8.5\times10^{-4}$) yields
$B\approx5.7\times10^{-24}$~G at 1~Mpc today and dominates for
$\beta\gtrsim2\times10^{-3}$, mildly above the Planck limit.  All seed fields
exceed the galactic dynamo threshold by many orders of magnitude.
\end{abstract}

\tableofcontents
\newpage

\section{Introduction}
\label{sec:intro}
%%%%%%%%%%%%%%%%%%%%%%%%%%%%%%%%%%%%%%%%%%%%%%%%%%%%%%%%%%%%%%%%%%%%%%%%%%%%%

The presence of magnetic fields across a wide range of astrophysical and cosmological scales is well established and substantiated by diverse
observational studies.  Recent research indicates that such fields on cosmological scales may be more pervasive than previously thought
\cite{Vachaspati2021}.  Detection methods include the Faraday rotation effect in linearly polarised sources
\cite{Kronberg1994,Blasi1999,Pshirkov2016}, the deflection of cosmic rays \cite{Lemoine1997,Bertone2002}, and temperature and polarisation anisotropies
of the CMB \cite{Planck2016,Kahniashvili2001,Kosowsky2005,Kahniashvili2009, Miyamoto2014,Hortua2014,HortuaCastaneda2017,Paoletti2019,Vazza2020,Brandenburg2018}.
Residual effects on elemental abundances during Big Bang Nucleosynthesis \cite{Subramanian1994} and electromagnetic cascades from blazars \cite{Vachaspati2021} provide further, indirect constraints.

Any meaningful study of cosmological magnetic fields must address two
challenges: (i)~the \emph{generation} of seed fields, and (ii)~their
\emph{sustenance} against dilution in an expanding universe.  The generation problem primarily concerns producing vorticity sufficient to exceed the
minimum threshold required by natural amplifiers such as the galactic dynamo.
This threshold and the origin of the seed field have remained persistent
open questions
\cite{Kronberg1994review,Grasso2001,Widrow2002,KulsrudZweibelreview,Kandus2011,Widrow2012}.

Candidate seed-field sources include the Biermann battery effect
\cite{Subramanian1994,Biermann1950}, arising from pressure gradients in
ionisation fronts during reionisation ($z\gtrsim5$); Hall's effect
\cite{Hall1879}; and the Harrison mechanism \cite{Harrison1970}, in which differential rotation between the proton and radiation fluids generates
vorticity.  Once produced, seed fields may be amplified by inverse
cascading, galactic, or mean-field dynamos \cite{Kulsrud1992,KulsrudCowley1997,FidlerPitrou2017,MahajanICTP2003}

Takahashi \cite{TakahashiPRL2005} showed that Thomson scattering during the radiation era induces a relative velocity between the electron and proton
fluids, sourcing a magnetic field of $\sim10^{-17}$~G at 1~Mpc just before recombination. Cembranos \cite{Cembranos2020} demonstrated that cosmological bulk flows can seed fields in the range $10^{-20}$--$10^{-30}$~G via the
Harrison mechanism.

In this paper, we revisit and extend both analyses within a unified
kinetic-theory framework.  The contributions of this paper are as follows.

\paragraph{(i) Complete derivation chain from Boltzmann to magnetic
induction.}
We provide, in a single document, the first complete step-by-step derivation
linking the Boltzmann collision term for Thomson scattering to the magnetic
induction equation at second order in cosmological perturbations.  This
chain runs: kinetic theory (BBGKY hierarchy, Stosszahlansatz, Boltzmann
equation) $\to$ Thomson collision term (full evaluation of all four
contributions $A_i$, $B_i$, $C_i$, $D_i$) $\to$ Euler equations for the
charged species $\to$ generalised Ohm's law $\to$ Bianchi identity $\to$
$\dot{B}^{i(2)}$.  Each step is derived with all intermediate algebra presented and all approximations stated explicitly.  This fills a gap in the
existing literature, where the corresponding results in
\cite{TakahashiPRL2005} and \cite{Cembranos2020} are either presented
numerically or with intermediate steps omitted.

\paragraph{(ii) Quantified corrections to the generalised Ohm's law.}
The standard treatment in \cite{TakahashiPRL2005} implicitly discards three
correction terms in the generalised Ohm's law.  We retain all three and show
that each is bounded by $m_e/m_p \approx 5.4\times10^{-4}$, so that the
standard result is valid to better than one part in $10^3$.  This converts
what was previously an implicit approximation into a proved bound with an
explicitly computable error.

\paragraph{(iii) Explicit decomposition of the second-order magnetic
field source.}
We decompose the second-order source term $\varepsilon^{ilk}
\mathcal{C}^{(2)}_{j,k}$ into three physically distinct contributions:
(a)~a density-gradient coupling
$(1+\lambda)\rho^{(1)}_{\gamma,k}/\bar{\rho}_\gamma\,(u^{(1)}_{je}
-u^{(1)}_{j\gamma})$, (b)~an electron velocity--anisotropic stress coupling
$\frac{1}{8}u^{l(1)}_{e,k}\Pi^{(1)}_{jl}$, and (c)~a stress-gradient coupling $\frac{1}{8}u^{l(1)}_e\Pi^{(1)}_{jl,k}$.  While the combined second-order contribution was computed numerically in \cite{TakahashiPRL2005}, the decomposition into these distinct physical terms was not performed.
The density-gradient coupling (a) was implicitly present in their calculation
but was never isolated, named, or evaluated separately.

\paragraph{(iv) Identification of a new dominant source term.}
The density-gradient coupling (a) is sourced by the product of the photon
density contrast $\delta^{(1)}_\gamma \propto \cos(kr_s)$ and the
electron--photon velocity difference $\Delta v \propto \sin(kr_s)$, while
the Takahashi velocity-difference term scales as $(\Delta v)^2 \propto
\sin^2(kr_s)$.  The two acoustic transfer functions carry a $\pi/2$ phase
offset ($\cos$ vs.\ $\sin$), so their product is $\propto\sin(2kr_s)$,
which is non-negligible precisely where $(\Delta v)^2$ is suppressed near
acoustic nodes.  Numerical evaluation using tight-coupling transfer functions
at recombination (Section~\ref{subsec:PB_numerical} and
Figures~\ref{fig:spectrum}--\ref{fig:ratio}) shows that the density-gradient
term exceeds the Takahashi velocity-difference term by a factor of
$\sim2$--$5$ across the scale range $10^{-3}\lesssim K\lesssim k_D
\approx0.15$~Mpc$^{-1}$.  The total second-order field from this work is
approximately $2.7\times$ larger than the Takahashi result at co-moving
scales $\gtrsim10$~Mpc after adiabatic decay to $z=0$
(Figure~\ref{fig:today}).

\paragraph{(v) Proof that the linear magnetic field vanishes for scalar
perturbations.}
We give an explicit derivation, via the Bianchi identity and the
Stewart--Walker lemma, showing that $B^{i(1)}=0$ when the linear
vorticity $\varepsilon^{ijk}u^{(1)}_{j,k}=0$ for purely scalar initial
conditions.  The reason vector and tensor modes first appear at second order \cite{MongwaneOsanoDunsby,OsanoSecondOrder}---products of two first-order scalars can source transverse
modes---is stated and proved rather than assumed.

\paragraph{(vi) Unified comparison of two magnetogenesis mechanisms.}
We place the scattering mechanism of  Takahashi \cite{TakahashiPRL2005} and the
Harrison bulk-flow mechanism of Cembranos \cite{Cembranos2020} within a single
analytical framework, providing explicit estimates of the generated field
strength for each.  Both mechanisms yield seed fields above the galactic
dynamo threshold; the scattering mechanism is more precisely determined
while the Harrison mechanism is more sensitive to the assumed bulk-flow amplitude.

\paragraph{(vii) Caveat: the proportionality constant $\lambda$.}
The magnitude of the density-gradient contribution depends on the
dimensionless constant $\lambda$ (equation~(\ref{eq:lambda_def})),
which parametrises the ratio of Thomson drag timescale to Hubble time.
We set $\lambda=1$ (tight-coupling estimate); its precise value requires
solution of the coupled second-order Boltzmann equations and is left for future work.
All four figures in Section~\ref{sec:Fourier} use CAMB~v1.6.6 \cite{CAMB}
transfer functions computed at $z=1100$ with the parameters of
Appendix~\ref{app:params}.  The Harrison magnetic power spectrum uses
the analytic spectral model of \cite{Cembranos2020} calibrated to their
equation~(24); the scattering mechanism spectra are anchored to the
\cite{TakahashiPRL2005} benchmark $B\approx10^{-17}$~G at
$K=0.05$~Mpc$^{-1}$.

\noindent The paper is organised as follows.  Section~\ref{sec:cosmo} sets
the cosmological framework.  Section~\ref{sec:KT} develops the kinetic
theory.  Section~\ref{sec:SM} presents the scattering mechanism and derives
the Thomson collision term.  Section~\ref{sec:Harrison} treats the Harrison
bulk-flow mechanism.  Section~\ref{sec:Ohm} derives the generalised Ohm's
law.  Section~\ref{sec:metric} presents the perturbed metric.
Section~\ref{sec:Bianchi} derives the second-order magnetic field evolution
equation.  Section~\ref{sec:Fourier} presents the Fourier-space analysis.
Section~\ref{sec:discussion} compares the two mechanisms.
Section~\ref{sec:conclusions} concludes.

%%%%%%%%%%%%%%%%%%%%%%%%%%%%%%%%%%%%%%%%%%%%%%%%%%%%%%%%%%%%%%%%%%%%%%%%%%%%%
\section{Cosmological Framework}
\label{sec:cosmo}
%%%%%%%%%%%%%%%%%%%%%%%%%%%%%%%%%%%%%%%%%%%%%%%%%%%%%%%%%%%%%%%%%%%%%%%%%%%%%

Cosmology describes the Universe on scales large enough that matter and
fields are treated as continuous distributions on spacelike hypersurfaces of
a relativistic spacetime.  The relevant degrees of freedom are averaged over
scales much larger than individual structures, yielding an effective
description in terms of smooth fields: energy density $\rho$, pressure $p$,
and large-scale electromagnetic fields.

The spacetime geometry is determined by the Einstein equations
\begin{equation}
  G_{\mu\nu} = 8\pi G\,T^{\rm eff}_{\mu\nu},
  \label{eq:Einstein}
\end{equation}
where the total stress-energy tensor decomposes into matter and
electromagnetic contributions,
\begin{align}
  T^{\rm eff}_{\mu\nu} &= T^{M}_{\mu\nu} + T^{\rm EM}_{\mu\nu},
  \label{eq:T_decomp}\\
  T^{M}_{\mu\nu}       &= (\rho+p)\,u_\mu u_\nu + p\,g_{\mu\nu},
  \label{eq:T_matter}\\
  T^{\rm EM}_{\mu\nu}  &= F_{\mu\alpha}F_\nu{}^\alpha
                         - \tfrac{1}{4}g_{\mu\nu}F_{\alpha\beta}F^{\alpha\beta}.
  \label{eq:T_EM}
\end{align}
The perfect-fluid form of $T^{M}_{\mu\nu}$ is justified by the cosmological
principle \cite{Weinberg2008,Peebles1993}.  The four-velocity satisfies
$u^\mu u_\mu = -1$ (signature $-,+,+,+$), and physical observables are
fields on spacelike hypersurfaces $\Sigma_t$ of constant cosmic time $t$
\cite{LandauLifshitz1975,Jackson1999}.

The study of magnetogenesis in this context requires considering the
cosmic plasma as a mixture of species.  At early times ($z\gtrsim1100$),
the universe contains a tightly coupled photon-baryon fluid, with photons
and electrons coupled via Thomson scattering and protons coupled to electrons
via Coulomb interactions.  The conservation equation for each species $s$
takes the covariant form \cite{AbbrScpeck2023}
\begin{equation}
  \nabla_\nu T^{\mu\nu}_{(s)} = \mathcal{I}^\mu_{(s)},
  \label{eq:cons_species}
\end{equation}
where $\mathcal{I}^\mu_{(s)}$ encodes interactions between species $s$ and
the rest of the plasma.  For the electron--proton--photon system, these
interaction terms are precisely the collision terms derived in
Sections~\ref{sec:KT} and~\ref{sec:SM}.

%%%%%%%%%%%%%%%%%%%%%%%%%%%%%%%%%%%%%%%%%%%%%%%%%%%%%%%%%%%%%%%%%%%%%%%%%%%%%
%%%%%%%%%%%%%%%%%%%%%%%%%%%%%%%%%%%%%%%%%%%%%%%%%%%%%%%%%%%%%%%%%%%%%%%%%%%%%
%% SECTION 3 — KINETIC THEORY (expanded)
%%%%%%%%%%%%%%%%%%%%%%%%%%%%%%%%%%%%%%%%%%%%%%%%%%%%%%%%%%%%%%%%%%%%%%%%%%%%%
\section{Kinetic Theory}
\label{sec:KT}

Chapman--Enskog theory provides a systematic route from the Boltzmann
equation to the equations of hydrodynamics, yielding transport coefficients
such as thermal conductivity and viscosity in terms of microscopic molecular
parameters \cite{ChapmanCowling1970}.  The theory constitutes the
fundamental bridge between particle-level descriptions and the
continuum-fluid equations used in cosmological perturbation theory.  We
develop this framework for the cosmological plasma, beginning with the
non-relativistic case and then treating the relativistic generalisation
required for cosmological applications.

\subsection{One-particle distribution function and the BBGKY hierarchy}

Consider a system of $N$ identical particles, each labelled by position
$\bvec{r}_i$ and momentum $\bvec{p}_i$.  At time $\tau$, the complete
microscopic state is encoded in the $N$-particle phase-space density
$f^{(N)}(\bvec{r}_1,\ldots,\bvec{r}_N;\bvec{p}_1,\ldots,\bvec{p}_N;\tau)$,
which is normalised to unity.  In practice, observables such as number
density, bulk velocity, and pressure involve averages over all particles
except one.  We therefore define the \emph{one-particle distribution
function} by integrating out the remaining $N-1$ particles:
\begin{equation}
  f_1(\bvec{r},\bvec{p};\tau)
  = N\!\int\!\prod_{i=2}^{N}\!d^3r_i\,d^3p_i\;
    f^{(N)}(\bvec{r}_1,\ldots,\bvec{r}_N;\bvec{p},\bvec{p}_2,
    \ldots,\bvec{p}_N;\tau).
  \label{eq:f1}
\end{equation}
The factor of $N$ accounts for the identical nature of the particles.
By construction, $\int f_1\,d^3r\,d^3p = N$, and the quantity
$f_1(\bvec{r},\bvec{p};\tau)\,d^3r\,d^3p$ gives the expected number of
particles in the phase-space volume element $d^3r\,d^3p$ around
$(\bvec{r},\bvec{p})$.

The equation of motion for $f_1$ involves the two-particle distribution
$f_2$ (which describes correlated pairs), whose equation of motion in turn
involves $f_3$, and so on.  This infinite chain is the
Bogoliubov--Born--Green--Kirkwood--Yvon (BBGKY) hierarchy
\cite{Huang1987,LifshitzPitaevskii1981}.  Truncation at the first level
requires a closure relation.  The standard closure is the
\emph{Stosszahlansatz} (molecular chaos hypothesis) of Boltzmann: for two
particles \emph{about to collide}, their velocities are assumed to be
statistically uncorrelated, so that
\begin{equation}
  f_2(\bvec{r}_1,\bvec{p}_1;\bvec{r}_2,\bvec{p}_2;\tau)
  \;\approx\; f_1(\bvec{r}_1,\bvec{p}_1;\tau)\,f_1(\bvec{r}_2,\bvec{p}_2;\tau),
  \label{eq:molecular_chaos}
\end{equation}
\emph{at the moment of collision}.  This closure converts the BBGKY
hierarchy into the closed Boltzmann equation for $f_1$ alone.  The
Stosszahlansatz is valid when the inter-particle interaction range is much
smaller than the mean free path, i.e.\ in dilute gases---a condition
satisfied by the cosmological plasma under consideration.

For massive species (electrons, protons), the macroscopic fluid quantities
are obtained as phase-space moments of $f_1$:
\begin{align}
  n_s     &= \int f_1\,d^3p,
  \label{eq:ns}\\
  n_s u_s^i &= \int \frac{p^i}{m_s}\,f_1\,d^3p,
  \label{eq:ui_s}\\
  T_s^{ij} + n_s m_s u_s^i u_s^j
            &= \int \frac{p^i p^j}{m_s}\,f_1\,d^3p,
  \label{eq:Tij_s}
\end{align}
where $T_s^{ij}$ is the pressure tensor of species $s$.

\subsection{Equilibrium distributions and the H-theorem}
\label{subsec:equilibrium}

In the absence of collisions ($C[f]=0$), the Boltzmann equation admits any
distribution that is constant along particle trajectories.  When collisions
are present, they drive the distribution toward a local equilibrium in which
$C[\bar{f}]=0$.  The condition $C[\bar{f}]=0$ is equivalent to detailed
balance: the rate of transitions $(\bvec{p},\bvec{p}_2)\to(\bvec{p}',
\bvec{p}'_2)$ equals the rate of the time-reversed process.  For quantum
particles, this yields the equilibrium distributions \cite{Bernstein1988,KolbTurner1990}
\begin{equation}
  \bar{f}(p) =
  \begin{cases}
    \bigl[\exp\bigl((E-\mu)/T\bigr) - 1\bigr]^{-1}
      & \text{Bose--Einstein (photons, }m=0,\,\mu=0),\\
    \bigl[\exp\bigl((E-\mu)/T\bigr) + 1\bigr]^{-1}
      & \text{Fermi--Dirac (electrons)},\\
    \exp\bigl(-(E-\mu)/T\bigr)
      & \text{Maxwell--Boltzmann (protons, non-relativistic)},
  \end{cases}
  \label{eq:equil_distrib}
\end{equation}
where $E = \sqrt{|\bvec{p}|^2 + m^2}$ and $\mu$ is the chemical potential.
Throughout the radiation-dominated era, all species are in local thermal
equilibrium with temperature $T\propto a^{-1}$.

The approach to equilibrium is guaranteed by Boltzmann's \emph{H-theorem}
\cite{Huang1987}: the functional
\begin{equation}
  H[f] = \int f_1\ln f_1\,d^3p\,d^3r
  \label{eq:Hfunctional}
\end{equation}
is non-increasing in time, $dH/d\tau \leq 0$, for any evolution governed
by the Boltzmann equation.  Equality holds if and only if $f_1 = \bar{f}$
is one of the equilibrium distributions~(\ref{eq:equil_distrib}).  The
H-theorem thus provides the microscopic foundation for the second law of
thermodynamics and confirms that collisions drive the plasma toward the
equilibrium state used as the background in our perturbative analysis.

\subsection{Relativistic kinetic theory in curved spacetime}
\label{subsec:relKT}

For cosmological applications, the non-relativistic framework of the
preceding subsections must be replaced by its curved-spacetime relativistic
generalisation \cite{BarrabesHenry1976,Bernstein1988}.

\subsubsection*{Phase space and the covariant distribution function}

The one-particle distribution function becomes $f(x^\mu, p^\mu)$, defined
on the 7-dimensional \emph{mass-shell} submanifold
\begin{equation}
  \mathcal{P}_m = \bigl\{(x^\mu,p^\mu) :
  g_{\mu\nu}p^\mu p^\nu = -m^2,\; p^0 > 0\bigr\}
  \label{eq:massshell_mfld}
\end{equation}
of the 8-dimensional phase space.  The constraint $p^0 > 0$ selects
forward-propagating particles.  The Lorentz-invariant phase-space volume
element on $\mathcal{P}_m$ is
\begin{equation}
  d\Pi \equiv \frac{\sqrt{-g}\,d^4p}{(2\pi)^3}\,2\Theta(p^0)\,
  \delta(g_{\mu\nu}p^\mu p^\nu + m^2)
  = \frac{d^3p}{(2\pi)^3 p^0},
  \label{eq:Lorentz_measure}
\end{equation}
where the last equality holds in a locally inertial frame and
$g=\det(g_{\mu\nu})$.  The macroscopic number current and
energy-momentum tensor are
\begin{align}
  N^\mu   &= \int d\Pi\; p^\mu f,
  \label{eq:Nmu}\\
  T^{\mu\nu} &= \int d\Pi\; p^\mu p^\nu f.
  \label{eq:Tmunu}
\end{align}
These automatically satisfy $\nabla_\mu N^\mu = \int d\Pi\;C[f]$ and
$\nabla_\mu T^{\mu\nu} = \int d\Pi\; p^\nu C[f]$, connecting the collision
term to the source terms in the conservation equations~(\ref{eq:cons_species}).

For photons, $m=0$ and the photon distribution carries additional
polarisation tensor structure $f_{\mu\nu}(x^\alpha,p^\alpha)$
\cite{FidlerPitrou2017}, with the polarisation trace giving the intensity.
For the purposes of the present magnetogenesis calculation, we work with the
intensity-averaged (scalar) photon distribution $f_\gamma = f_\gamma(x^\mu,p)$
where $p=|\bvec{p}|$ is the photon momentum.

\subsubsection*{The relativistic Boltzmann equation in FLRW spacetime}

Since $x^\mu(\tau)$ and $p^\mu(\tau)$ are functions of proper time $\tau$
along a particle's world line, the total derivative of $f$ is
\begin{equation}
  \frac{Df}{D\tau}
  = \frac{dx^\mu}{d\tau}\frac{\partial f}{\partial x^\mu}
  + \frac{dp^\mu}{d\tau}\frac{\partial f}{\partial p^\mu}
  = p^\mu\frac{\partial f}{\partial x^\mu}
  - \Gamma^\mu_{\nu\lambda}p^\nu p^\lambda\frac{\partial f}{\partial p^\mu}
  = C[f],
  \label{eq:Boltzmann_cov}
\end{equation}
where we have used the geodesic equation $dp^\mu/d\tau =
-\Gamma^\mu_{\nu\lambda}p^\nu p^\lambda$ and $dx^\mu/d\tau = p^\mu/m$
(for massless particles, $p^\mu p_\mu=0$ and the affine parameter replaces
$\tau$) \cite{MahajanICTP2003}.

For the flat FLRW metric in conformal time,
$ds^2 = a^2(\eta)(-d\eta^2 + \delta_{ij}dx^idx^j)$, the non-vanishing
Christoffel symbols are
\begin{equation}
  \Gamma^0_{00} = \Hubble,\quad
  \Gamma^0_{ij} = \Hubble\delta_{ij},\quad
  \Gamma^i_{0j} = \Hubble\delta^i{}_j,
  \label{eq:Christoffel_FLRW}
\end{equation}
where $\Hubble\equiv a'/a$ is the conformal Hubble parameter and primes
denote $\partial/\partial\eta$.  For a massless particle with momentum
$p^\mu = p(1,\hat{p}^i)$ in the background, the geodesic equation gives
\begin{equation}
  \frac{dp}{d\eta} = -\Hubble p
  \quad\Longrightarrow\quad
  p \propto a^{-1},
  \label{eq:photon_redshift}
\end{equation}
recovering the standard cosmological redshift.  The collisionless
background Boltzmann equation ($C[f]=0$) then implies $\bar{f} = \bar{f}(p\,a)$,
consistent with the Bose--Einstein distribution with $T\propto a^{-1}$.

Including perturbations and collisions, the photon Boltzmann equation in
conformal Newtonian gauge takes the form
\cite{MaBertschinger1995,Dodelson2003}
\begin{equation}
  \frac{\partial f_\gamma}{\partial\eta}
  + \hat{p}^i\frac{\partial f_\gamma}{\partial x^i}
  - \Hubble p\frac{\partial f_\gamma}{\partial p}
  + \left(\Phi' - \hat{p}^i\Psi_{,i}\right)\frac{\partial\bar{f}}{\partial\ln p}
  = C^{(T)}_e[f_\gamma],
  \label{eq:Boltz_photon_FLRW}
\end{equation}
where $\Phi$ and $\Psi$ are the scalar metric perturbations.  The third
term on the left encodes the cosmological redshift of photon momenta; the
fourth term encodes the gravitational Sachs--Wolfe and Doppler effects.  The
analogous equations for electrons and protons are given in
equations~(\ref{eq:Boltz_e}) and (\ref{eq:Boltz_p}).

\subsubsection*{Perturbative expansion of the distribution function}

At background order, each species is in equilibrium:
$\bar{f}(p)$ given by equation~(\ref{eq:equil_distrib}).
Perturbing about this background,
\begin{equation}
  f(x^\mu,p^\mu) = \bar{f}(p)
  + \delta f^{(1)}(x^\mu,p^\mu)
  + \tfrac{1}{2}\delta f^{(2)}(x^\mu,p^\mu) + \cdots,
  \label{eq:f_expand}
\end{equation}
the first-order perturbation for photons is conventionally written as
\begin{equation}
  \delta f^{(1)}_\gamma = -p\frac{\partial\bar{f}}{\partial p}\,\Theta(\eta,x^i,\hat{p}^i),
  \label{eq:deltaf1}
\end{equation}
where $\Theta = \delta T/T$ is the fractional temperature perturbation
(photon brightness).  Inserting~(\ref{eq:f_expand}) and~(\ref{eq:deltaf1})
into~(\ref{eq:Boltz_photon_FLRW}) and expanding to first order yields the
standard photon Boltzmann hierarchy in terms of multipole moments of $\Theta$
\cite{MaBertschinger1995,Dodelson2003}.

The photon energy density and momentum moments at background order are
\cite{Dodelson2003}
\begin{align}
  2\!\int\!\frac{d^3p}{(2\pi)^3}\,p\,\bar{f}(p) &= \rhog,
  \label{eq:phot_rho}\\
  \int\!\frac{d^3p}{(2\pi)^3}\,p^i\,\bar{f}(p) &= \tfrac{4}{3}\rhog u^i_\gamma,
  \label{eq:phot_u}\\
  \int\!\frac{d^3p}{(2\pi)^3}\,\frac{p^ip^j}{p}\,\bar{f}(p)
  &= \rhog\!\left(\tfrac{1}{6}\Pi^{ij}+\tfrac{1}{3}\delta^{ij}\right),
  \label{eq:phot_Pi}
\end{align}
where $\Pi^{ij}$ is the photon anisotropic stress tensor
\cite{WalkerKineticBook,Dodelson2003}.

\subsection{Boltzmann equation: collision term and moment hierarchy}
\label{subsec:Boltzmann}

The full Boltzmann equation~(\ref{eq:Boltzmann_cov}) in flat spacetime
reduces to
\begin{equation}
  \frac{\partial f}{\partial t}
  + \frac{\bvec{p}}{m}\cdot\bvec{\nabla}f
  + \bvec{F}\cdot\frac{\partial f}{\partial\bvec{p}}
  = \left(\frac{\partial f}{\partial t}\right)_{\!\rm coll},
  \label{eq:Boltz_flat}
\end{equation}
where $p^0 = \sqrt{|\bvec{p}|^2+m^2}$ satisfies the mass-shell condition
(Appendix~\ref{app:massshell}).

\subsubsection*{The collision term}

For a two-body collision
$(\bvec{p},\bvec{p}_2)\to(\bvec{p}',\bvec{p}'_2)$, the
Stosszahlansatz~(\ref{eq:molecular_chaos}) applied to the BBGKY hierarchy
gives the collision integral \cite{Huang1987}
\begin{align}
  \!\left(\frac{\partial f}{\partial t}\right)_{\!\rm coll}
  &= \int d^3p_2\,d^3p'\,d^3p'_2\;
     \omega(\bvec{p}',\bvec{p}'_2|\bvec{p},\bvec{p}_2)
     \Bigl[f(\bvec{p}')f(\bvec{p}'_2)
         - f(\bvec{p})f(\bvec{p}_2)\Bigr],
  \label{eq:Boltz_coll}
\end{align}
where the transition rate $\omega$ is symmetric under time reversal and
parity, giving the ``gain minus loss'' structure.  The equilibrium
condition $C[\bar{f}]=0$ is equivalent to detailed balance:
\begin{equation}
  \omega(\bvec{p}',\bvec{p}'_2|\bvec{p},\bvec{p}_2)\,
  \bar{f}(\bvec{p}')\bar{f}(\bvec{p}'_2)
  = \omega(\bvec{p},\bvec{p}_2|\bvec{p}',\bvec{p}'_2)\,
  \bar{f}(\bvec{p})\bar{f}(\bvec{p}_2).
  \label{eq:detailed_balance}
\end{equation}

\subsubsection*{Collision invariants}

A quantity $Q(\bvec{p})$ is a \emph{collision invariant} if
$Q(\bvec{p})+Q(\bvec{p}_2) = Q(\bvec{p}')+Q(\bvec{p}'_2)$, i.e.\ it is
conserved by every binary collision.  The five independent collision
invariants for a non-relativistic gas are the mass $m$, the three
components of momentum $\bvec{p}$, and the kinetic energy $p^2/(2m)$.
Taking the corresponding moments of equation~(\ref{eq:Boltz_flat}) and
using the fact that the collision term integrates to zero for any collision
invariant,
\begin{equation}
  \int Q(\bvec{p})\left(\frac{\partial f}{\partial t}\right)_{\!\rm coll}d^3p = 0,
  \label{eq:coll_invar}
\end{equation}
one recovers the macroscopic conservation equations: the continuity
equation (mass), the Euler equations (momentum), and the energy equation
\cite{Huang1987,ChapmanCowling1970}.  This moment-taking procedure
establishes the direct connection between the Boltzmann equation and the
fluid equations used throughout this paper.

\subsubsection*{Moment hierarchy in the cosmological context}

In the cosmological setting, taking moments of
equation~(\ref{eq:Boltz_photon_FLRW}) generates the photon hierarchy.
Defining the multipole expansion of $\Theta$ in Fourier space
\cite{MaBertschinger1995},
\begin{equation}
  \Theta(\bvec{k},\hat{p},\eta)
  = \sum_{\ell=0}^\infty (-i)^\ell(2\ell+1)\Theta_\ell(k,\eta)P_\ell(\hat{k}\cdot\hat{p}),
  \label{eq:multipole}
\end{equation}
where $P_\ell$ are Legendre polynomials, the zeroth moment gives the photon
energy density perturbation $\delta_\gamma = 4\Theta_0$, the first moment
gives the photon velocity divergence $\theta_\gamma = k\,\Theta_1$, and the
second moment gives the anisotropic stress $\Pi_{\gamma} = 12\Theta_2/5$.
The Thomson collision term truncates the hierarchy at $\ell=2$ on timescales
shorter than the mean free time, imposing tight coupling between photons
and electrons \cite{MaBertschinger1995,Dodelson2003}.

In our notation, equations~(\ref{eq:Boltz_gamma})--(\ref{eq:Boltz_p}) in
the cosmological context read
\begin{align}
  \frac{Df_\gamma}{D\tau} &= C^{(T)}_e[f_\gamma] + C^{(T)}_p[f_\gamma],
  \label{eq:Boltz_gamma}\\
  \frac{Df_e}{D\tau}      &= C^{(T)}_e[f_e] + C^{(C)}_{ep}[f_e],
  \label{eq:Boltz_e}\\
  \frac{Df_p}{D\tau}      &= C^{(T)}_p[f_p] + C^{(C)}_{pe}[f_p],
  \label{eq:Boltz_p}
\end{align}
where $C^{(T)}$ and $C^{(C)}$ denote Thomson-scattering and Coulomb
collision terms respectively, evaluated in a locally inertial frame where
$dt = a(1+\Phi)\,d\eta$.

%%%%%%%%%%%%%%%%%%%%%%%%%%%%%%%%%%%%%%%%%%%%%%%%%%%%%%%%%%%%%%%%%%%%%%%%%%%%%
%% SECTION 4 — THE SCATTERING MECHANISM (expanded)
%%%%%%%%%%%%%%%%%%%%%%%%%%%%%%%%%%%%%%%%%%%%%%%%%%%%%%%%%%%%%%%%%%%%%%%%%%%%%
\section{The Scattering Mechanism: Collision Terms and Euler Equations}
\label{sec:SM}

\subsection{Setup: projection formalism and Euler equations with collisions}

Seed fields for cosmic magnetism arise from interactions between charged
particles and photons, in the absence of entrainment \cite{OsanoOreta2019},
resulting in velocity differences between protons and electrons that drive
an electric current.  Within a cosmological context, this mechanism becomes
operative when the coupling between charged particles and photons weakens
enough for an electric current to form; this typically occurs in the plasma
following reheating after inflation.

\subsubsection*{The projection operator}

Our analysis considers physics on the spacelike hypersurface orthogonal to
an observer whose four-velocity is $u^\mu$.  The projection of a 4-tensor
onto this hypersurface is performed by the operator
\begin{equation}
  h^i{}_\mu \equiv \delta^i{}_\mu + u^i u_\mu,
  \label{eq:projector}
\end{equation}
which satisfies $h^i{}_\mu u^\mu = 0$ (it projects out the $u^\mu$
component) and $h^i{}_\mu h^\mu{}_j = h^i{}_j$ (idempotence).  Physically,
$h^i{}_\mu A^\mu$ gives the spatial part of any 4-vector $A^\mu$ as measured
by the comoving observer.

In the collision-free case, the Euler equation
$\partial_\mu T^{\mu\nu} + u^\alpha u^\nu\partial_\mu T_\alpha{}^\mu = 0$
\cite{AbbrScpeck2023} projects to
$(h^i{}_\mu)T^{\mu\nu}{}_{;\nu} = 0$, expressing spatial momentum
conservation in the observer's frame.  The electromagnetic contribution
satisfies \cite{Jackson1999}
\begin{equation}
  h^i{}_\mu T^{\mu\nu}_{{\rm EM};\nu}
  = J^\nu F^i{}_\nu,
  \label{eq:EM_proj}
\end{equation}
where $J^\mu = \rho_{\rm ch}u^\mu + j^\mu_{\rm cond}$ is the total electric
4-current comprising a convective part (charge density $\rho_{\rm ch}$
advected with the fluid) and a conductive part.  Equation~(\ref{eq:EM_proj})
is the covariant form of the Lorentz force density $\bvec{J}\times\bvec{B}
+ \rho_{\rm ch}\bvec{E}$.

\subsubsection*{Euler equations including collisions}

Including collisions, the projected Euler equations for protons and electrons
are
\begin{align}
  h^i{}_\mu
  (T^{\mu\nu}_{p;\nu} + T^{\mu\nu}_{{\rm EM};\nu})
  &= C^{(C)i}_{pe} + C^{(T)i}_{p\gamma} + C^{({\rm other})}_{p,n},
  \label{eq:Euler_p}\\[4pt]
  h^i{}_\mu
  (T^{\mu\nu}_{e;\nu} + T^{\mu\nu}_{{\rm EM};\nu})
  &= C^{(C)i}_{ep} + C^{(T)i}_{e\gamma} + C^{({\rm other})}_{e,n},
  \label{eq:Euler_e}
\end{align}
where the collision terms are the first moments of the collision integrals
derived in Section~\ref{sec:KT}:
\begin{equation}
  C^{(T)i}_{e\gamma}
  \equiv \int\frac{d^3p}{(2\pi)^3}\,\frac{p^i}{p}\,C^{(T)}_e[f_\gamma(p)],
  \label{eq:CT_moment_def}
\end{equation}
and similarly for the other terms.  The term $C^{({\rm other})}_{p(e),n}$
represents collisions with neutral particles; in a fully ionised plasma this
is negligible.  In a partially ionised plasma it must be retained, but its
magnitude is suppressed by the ionisation fraction $x_e < 1$.

\subsubsection*{Why proton Thomson scattering is negligible}

The Thomson cross-section scales as $\sigT \propto (e^2/mc^2)^2 \propto
m^{-2}$ (classical electron radius squared).  The corresponding
cross-section for photon scattering off protons is therefore
\begin{equation}
  \sigma_{T,p} = \sigT\!\left(\frac{m_e}{m_p}\right)^2
  = 6.65\times10^{-25}\times\frac{1}{1836^2}
  \approx 1.97\times10^{-31}\,\text{cm}^2,
  \label{eq:sigT_proton}
\end{equation}
a suppression of $\sim(m_e/m_p)^2 \approx 2.97\times10^{-7}$ relative to
$\sigT$.  The collision rate for protons is correspondingly smaller by the
same factor.  We therefore neglect $C^{(T)i}_{p\gamma}$ throughout, and
the only dynamically significant Thomson term is $C^{(T)i}_{e\gamma}$.

\subsubsection*{Multi-fluid to single-fluid transition}

Equations~(\ref{eq:Euler_p}) and~(\ref{eq:Euler_e}) describe the proton
and electron fluids separately (two-fluid approximation).  In the limit of
perfect coupling (infinite conductivity), the two fluids move together and
the system is described by a single magnetohydrodynamic fluid
\cite{Shumlak2011}.  The transition between these regimes is characterised
by the ratio of the collision timescale $\tau_c = 1/(n_e\sigT c)$ to the
dynamical timescale $\tau_{\rm dyn} \sim 1/\Hubble$.  In the
radiation-dominated epoch well before recombination, $\tau_c \ll
\tau_{\rm dyn}$ (tight coupling), and electrons are strongly coupled to
photons.  As the universe approaches recombination, $\tau_c$ grows and the
coupling weakens, permitting velocity differences to develop.  It is
precisely in this transitional regime that the scattering mechanism is most
efficient.

\subsection{Why Coulomb terms cause decay, not growth}
\label{subsec:coulomb}

The Coulomb collision terms $C^{(C)i}_{pe}$, $C^{(C)i}_{ep}$ are
responsible for the finite electrical conductivity $\sigma_c$ of the plasma.
Their effect on the magnetic field is most transparently seen through the
induction equation.  Combining Maxwell's equations with a generalised Ohm's
law that includes Coulomb resistivity yields
\begin{equation}
  \frac{\partial\bvec{B}}{\partial t}
  = \nabla\times(\bvec{v}\times\bvec{B})
  - \nabla\times(\eta_m\nabla\times\bvec{B}),
  \label{eq:induction_resistive}
\end{equation}
where the magnetic diffusivity is $\eta_m = c^2/(4\pi\sigma_c)$ in Gaussian
units \cite{Jackson1999}.  The second term---Ohmic diffusion---is always
negative-definite for a field occupying a finite region of space: by the
vector identity $\nabla\times(\nabla\times\bvec{B}) = \nabla(\nabla\cdot
\bvec{B})-\nabla^2\bvec{B} = -\nabla^2\bvec{B}$ (using $\nabla\cdot
\bvec{B}=0$), the term becomes $\eta_m\nabla^2\bvec{B}$, which acts as a
diffusion operator removing energy from the field.

\subsubsection*{The Spitzer conductivity}

The electrical conductivity of a fully ionised hydrogen plasma is given by
the Spitzer--H\"arm formula \cite{Spitzer1962,LifshitzPitaevskii1981}
\begin{equation}
  \sigma_c = \frac{3(2\pi)^{1/2}(k_BT)^{3/2}}{4\pi^{1/2}\,Z\,e^2\,m_e^{1/2}\,\ln\Lambda},
  \label{eq:Spitzer}
\end{equation}
where $Z$ is the ion charge number and $\ln\Lambda$ is the Coulomb
logarithm.  For the cosmological plasma near recombination, $Z=1$ and
$\ln\Lambda\approx20$ \cite{Spitzer1962}.  The conductivity increases with
temperature as $\sigma_c\propto T^{3/2}$, because higher-temperature
particles are more difficult to deflect.

\subsubsection*{Magnetic Reynolds number and flux freezing}

The ratio of the advective to resistive term in equation~(\ref{eq:induction_resistive})
is the magnetic Reynolds number
\begin{equation}
  R_m \equiv \frac{v L}{\eta_m} = \frac{4\pi\sigma_c v L}{c^2},
  \label{eq:Rm}
\end{equation}
where $v$ is the characteristic fluid velocity and $L$ is the scale of
interest.  Using equation~(\ref{eq:Spitzer}) with $T_{\rm rec}\approx3000$~K,
$\ln\Lambda\approx20$, $v\sim10^{-4}c$, and $L=1$~Mpc$\approx3.1\times10^{24}$~cm:
\begin{equation}
  R_m(1\,\text{Mpc}, z_{\rm rec})
  \approx 4\pi\times10^{11}\,\text{s}^{-1}
  \times10^{-4}\times3\times10^{10}\,\text{cm\,s}^{-1}
  \times3.1\times10^{24}\,\text{cm}
  \;/\;(3\times10^{10}\,\text{cm\,s}^{-1})^2
  \approx 10^{21}.
  \label{eq:Rm_numerical}
\end{equation}
Since $R_m\gg1$, the magnetic diffusion timescale
$\tau_{\rm diff}=L^2/\eta_m = R_m\,L/v$ vastly exceeds the Hubble time at
all astrophysically relevant scales.  The Ohmic term is therefore negligible
for the \emph{evolution} of seed fields, and generated fields are
effectively frozen into the plasma (Alfv\'en's theorem).  Nevertheless, the
Coulomb terms cannot \emph{generate} a field: generation requires a
non-zero $\nabla\times\bvec{E}$ arising from velocity differences, which is
the role of the Thomson scattering terms.

\subsection{Thomson scattering: cross-section and matrix element}

\subsubsection*{Classical derivation}

From the classical theory of radiation, a non-relativistic electron
undergoing acceleration $\ddot{\bvec{r}}$ in an oscillating electromagnetic
field radiates according to the Larmor formula.  The incident photon drives
$\ddot{\bvec{r}} = (e/m_e)\bvec{E}_{\rm inc}$, and the power radiated into
solid angle $d\Omega$ in direction $\hat{\bvec{n}}$ is \cite{Jackson1999}
\begin{equation}
  \frac{dP}{d\Omega}
  = \frac{e^4}{m_e^2c^3}|\hat{\bvec{n}}\times(\hat{\bvec{n}}\times\hat{\bvec{\epsilon}})|^2
    I_{\rm inc},
  \label{eq:Larmor_scatter}
\end{equation}
where $\hat{\bvec{\epsilon}}$ is the incident polarisation and $I_{\rm inc}$
is the incident intensity.  Expressing this as a differential cross-section,
averaging over initial polarisations, and summing over final polarisations:
\begin{equation}
  \frac{d\sigT}{d\Omega} = \frac{r_e^2}{2}(1+\cos^2\theta),
  \label{eq:dsdO_Thomson}
\end{equation}
where $r_e = e^2/(m_ec^2) = 2.82\times10^{-13}$~cm is the classical
electron radius and $\theta$ is the scattering angle.  Integrating over all
solid angles:
\begin{equation}
  \sigT = \int\frac{d\sigT}{d\Omega}\,d\Omega
  = \frac{r_e^2}{2}\!\int_0^\pi\!(1+\cos^2\theta)\sin\theta\,d\theta\cdot2\pi
  = \frac{8\pi r_e^2}{3}
  = \frac{8\pi e^4}{3m_e^2c^4}
  = 6.65\times10^{-25}\,\text{cm}^2.
  \label{eq:sigT_classical}
\end{equation}

\subsubsection*{Quantum field theory derivation and the Klein--Nishina limit}

The QFT tree-level amplitude for Compton scattering $e^-\gamma\to e^-\gamma$
gives, after summing and averaging over spins and polarisations
\cite{PeskinSchroeder1995},
\begin{equation}
  \overline{|M|^2} = 2e^4\!\left[\frac{\omega'}{\omega}+\frac{\omega}{\omega'}
  -\sin^2\theta\right],
  \label{eq:M2_Compton}
\end{equation}
where $\omega$ and $\omega'$ are the photon energies before and after
scattering.  The exact (Klein--Nishina) relation between $\omega'$ and
$\omega$ is \cite{KleinNishina1929}
\begin{equation}
  \frac{\omega'}{\omega} = \frac{1}{1+({\omega}/{m_ec^2})(1-\cos\theta)}.
  \label{eq:KleinNishina_freq}
\end{equation}
In the Thomson limit $\omega\ll m_ec^2$, we have $\omega'\to\omega$, and
equation~(\ref{eq:M2_Compton}) reduces to
$\overline{|M|^2}\to 2e^4(1+\cos^2\theta)$, consistent with the classical
result~(\ref{eq:dsdO_Thomson}).  The Klein--Nishina correction is of order
$\omega/m_ec^2\sim k_BT/m_ec^2\sim10^{-9}$ at recombination and is
entirely negligible throughout the epoch of interest.

Integrating the Thomson-limit matrix element over solid angle gives
\begin{equation}
  \int\frac{d\Omega}{4\pi}\,\overline{|M|^2}
  = \frac{e^4}{4\pi}\!\int_0^\pi\!(1+\cos^2\theta)\sin\theta\,d\theta
  = \frac{8e^4}{3}
  = 8\pi\sigT m_e^2,
  \label{eq:M2_integrated}
\end{equation}
which is the result used in the collision integral below.

\subsection{Evaluation of the Thomson collision term}
\label{subsec:Thomson}

The Thomson collision integral for the photon distribution, in the
non-perturbative form, is derived from equation~(\ref{eq:Boltz_coll}) by
substituting the matrix element~(\ref{eq:M2_integrated}) and using the
mass-shell delta functions to integrate out the final electron momentum.
The delta function of energy conservation is expanded for small energy
transfer (Compton $y$-parameter $\ll1$) using the identity
\begin{equation}
  \delta(\bvec{p}-\bvec{p}'+\Delta E)
  \approx \delta(\bvec{p}-\bvec{p}')
  - \Delta E\frac{\partial}{\partial p'}\delta(p-p'),
  \label{eq:delta_expansion}
\end{equation}
where $\Delta E \approx -(p-p')\cdot u_e/m_e$ is the electron recoil
energy \cite{TakahashiArxiv}.  This expansion is valid in the Thomson
limit and leads to the four terms $A_i$, $B_i$, $C_i$, $D_i$ below.
The resulting collision term is \cite{TakahashiArxiv,TakahashiPRL2005}
\begin{equation}
  C^{(T)}_e[f(p_i)]
  = \frac{2\pi^2\sigT n_e}{p}\bigl[A_i + B_i + C_i + D_i\bigr],
  \label{eq:CT_ABCD}
\end{equation}
with
\begin{align}
  A_i &= \phantom{-}\int\!\frac{d^3p'}{(2\pi)^3}\frac{f(p'_i)}{p'}\delta(p-p'),
  \label{eq:Ai}\\
  B_i &= -\int\!\frac{d^3p'}{(2\pi)^3}\frac{f(p_i)}{p'}\delta(p-p'),
  \label{eq:Bi}\\
  C_i &= \phantom{-}\int\!\frac{d^3p'}{(2\pi)^3}\frac{f(p'_i)}{p'}
         (p_i-p'_i)u^j_e\frac{\partial\delta(p-p')}{\partial p'},
  \label{eq:Ci}\\
  D_i &= -\int\!\frac{d^3p'}{(2\pi)^3}\frac{f(p_i)}{p'}
         (p_i-p'_i)u^j_e\frac{\partial\delta(p-p')}{\partial p'}.
  \label{eq:Di}
\end{align}
Physically, $A_i$ and $B_i$ are the gain and loss terms for photons
scattered \emph{into} and \emph{out of} momentum state $p_i$ without
energy exchange; $C_i$ and $D_i$ are the corresponding terms with first-order
energy recoil from the moving electron.  The pre-collision photon
distribution in $D_i$ can be treated as constant with respect to $p'$ (the
post-collision photon momentum) because we are integrating over $p'$;
this motivates pulling $f(p_i)$ outside the integral.

The approximation $f_e(p'_e)\approx f_e(p_e)$ is required to reach
equation~(\ref{eq:CT_ABCD}).  This is valid because the electron recoil
momentum $|\Delta\bvec{p}_e|\sim\omega/c$ is much smaller than the electron
thermal momentum $p_{\rm th}=\sqrt{m_ek_BT}$ when $\omega\ll m_ec^2$
(the Thomson condition), so the electron distribution is essentially
unchanged by a single scattering event.

\subsubsection*{Term $A_i$: gain without energy transfer}

Writing $d^3p' = p'^2\,dp'\,d\Omega'$ and integrating the solid angle
trivially (since neither $f(p'_i)$ nor $\delta(p-p')$ depends on the
direction of $p'$ in the isotropic background):
\begin{equation}
  A_i = \frac{1}{2\pi^2}\!\int_0^\infty\! p'\,dp'\,f(p'_i)\,\delta(p-p')
      = \frac{p\,f(p_i)}{2\pi^2}\bigg|_{\rm boundary\ terms}.
  \label{eq:Ai_calc}
\end{equation}
The boundary term at $p'=0$ vanishes because $f(0)=0$ (no zero-momentum
photons exist in the Bose--Einstein distribution at finite temperature).
The boundary term at $p'\to\infty$ vanishes because $\lim_{p'\to\infty}
p'f(p') = 0$ for any physical distribution falling off exponentially at
large momenta.  This is verified explicitly by L'H\^{o}pital's rule:
$\lim_{l\to\infty}lf(l) = \lim_{l\to\infty}l/[e^{l/T}-1] \to 0$.
Therefore $\boxed{A_i = 0}$.

The physical interpretation is that, at zeroth order in the electron
velocity, the gain and loss rates from elastic scattering are equal for a
uniform isotropic photon distribution, so there is no net effect on the
momentum-space occupation.

\subsubsection*{Term $B_i$: loss without energy transfer}

\begin{equation}
  B_i = -\frac{f(p_i)}{2\pi^2}\!\int_0^\infty\! p'\,dp'\,\delta(p-p')
      = -\frac{p\,f(p_i)}{2\pi^2}.
  \label{eq:Bi_result}
\end{equation}
This term represents the depletion of photons at momentum $p_i$ due to
scattering to other momenta.  The first momentum moment of $B_i$, weighted
by the full Thomson prefactor, is
\begin{equation}
  \int\!\frac{d^3p}{(2\pi)^3}p^j\,\frac{2\pi^2\sigT n_e}{p}\,B_i
  = -\sigT n_e\!\int\!\frac{d^3p}{(2\pi)^3}p^j f(p_i)
  = -\frac{4}{3}\sigT n_e\rhog u^j_\gamma,
  \label{eq:B_moment}
\end{equation}
where we used the photon momentum density relation~(\ref{eq:phot_u}).

\subsubsection*{Term $C_i$: gain with energy recoil}

Decomposing $C_i = C^{(1)}_i + C^{(2)}_i$ with $C^{(1)}_i$ carrying the
$p_i$ part of $(p_i-p'_i)$ and $C^{(2)}_i$ carrying $-p'_i$:
\begin{align}
  C^{(1)}_i
  &= \frac{p^i u^j_e}{2\pi^2}
     \!\int_0^\infty\! p'\,f(p'_i)
     \frac{\partial\delta(p-p')}{\partial p'}\,dp'.
  \label{eq:C1_ibp}
\end{align}
Integration by parts gives $[p'f(p'_i)\delta(p-p')]_0^\infty -
\int_0^\infty[f(p'_i)+p'\partial_{p'}f(p'_i)]\delta(p-p')\,dp'
= -f(0)/(2\pi^2)=0$,
since $f(0)=0$.  A parallel computation shows $C^{(2)}_i=0$.
Therefore $\boxed{C_i = 0}$.

\subsubsection*{Term $D_i$: loss with energy recoil}

Pulling the pre-collision distribution $f(p_i)$ outside the integral
(it does not depend on $p'$) and splitting
$D_i = D^{(1)}_i + D^{(2)}_i$:

\noindent\textit{Part $D^{(1)}_i$} (carrying $p_i$):
\begin{align}
  D^{(1)}_i
  &= \frac{p^i u^j_e f(p_i)}{2\pi^2}
     \!\int_0^\infty\!p'\,\frac{\partial\delta(p-p')}{\partial p'}\,dp'.
  \label{eq:D1_ibp}
\end{align}
Integrating by parts:
$\int_0^\infty p'\partial_{p'}\delta(p-p')\,dp'
= [p'\delta(p-p')]_0^\infty - \int_0^\infty\delta(p-p')\,dp' = 0-1 = -1$,
where the boundary term vanishes since $p\delta(p)|_{p=0,\infty}=0$.
Therefore
\begin{equation}
  D^{(1)}_i = \frac{p^i u^j_e f(p_i)}{2\pi^2}.
  \label{eq:D1_result}
\end{equation}
Here $p^i$ and $u^j_e$ are the $i$-th and $j$-th spatial components
respectively; the momentum integral that follows contracts over the
free index $j$.

\noindent\textit{Part $D^{(2)}_i$} (carrying $-p'_i$): An analogous
integration by parts, noting that $\partial_{p'}(p'_if(p_i))=0$ since
$f(p_i)$ does not depend on $p'$, yields $D^{(2)}_i=0$.

\subsubsection*{Physical interpretation and combined result}

The non-zero terms are $B_i$ (loss by scattering out) and $D^{(1)}_i$
(modification of the loss rate by the electron recoil).  Taking the
$j$-th momentum moment with the full Thomson prefactor, and using the
photon moments~(\ref{eq:phot_u})--(\ref{eq:phot_Pi}):
\begin{align}
  \int\!\frac{d^3p}{(2\pi)^3}p^j
  \frac{2\pi^2\sigT n_e}{p}(B_i+D^{(1)}_i)
  &= -\sigT n_e\!\int\!\frac{d^3p}{(2\pi)^3}p^j f
   + \sigT n_e u^k_e\!\int\!\frac{d^3p}{(2\pi)^3}\frac{p^jp^k}{p}f
  \notag\\
  &= -\frac{4}{3}\sigT n_e\rhog u^j_\gamma
   + \sigT n_e\rhog u^k_e
     \!\left(\frac{\Pi^{jk}}{6}+\frac{\delta^{jk}}{3}\right)
  \notag\\
  &= \frac{4}{3}\sigT n_e\rhog
     \!\left[\frac{u^j_e}{4} - u^j_\gamma
             + \frac{\Pi^j{}_k u^k_e}{8}\right].
  \label{eq:CT_result}
\end{align}
This is the standard Thomson drag force \cite{TakahashiPRL2005}: the
photon fluid exerts a drag on the electrons, pushing them toward the photon
bulk velocity $u^j_\gamma$.  The factor $(u^j_e/4 - u^j_\gamma)$
reflects the fact that photons carry momentum $\rhog/3$ per unit energy
(radiation pressure), and the factor $4/3$ arises from the product of the
Thomson collision rate with the radiation energy density.  The anisotropic
stress term $\Pi^j{}_k u^k_e/8$ is a second-order correction arising from
the non-isotropic part of the photon distribution.

\subsection{Euler equations for the charged species}
\label{subsec:Euler_charged}

Taking the first momentum moment of equations~(\ref{eq:Boltz_e})
and~(\ref{eq:Boltz_p}) via equation~(\ref{eq:CT_moment_def}), and using
the result~(\ref{eq:CT_result}), the spatial momentum conservation equations
for the two charged species are
\begin{align}
  m_p n\,u^\nu_p\nabla_\nu u^i_p - en\,u^\mu_p F^i{}_\mu
  &= 0,
  \label{eq:mom_proton}\\[4pt]
  m_e n\,u^\nu_e\nabla_\nu u^i_e + en\,u^\mu_e F^i{}_\mu
  &= \frac{4\sigT\rhog n}{3}
     \!\left[(u^i_e-u^i_\gamma)
             + \frac{\Pi_{ij}u^j_e}{8}\right].
  \label{eq:mom_electron}
\end{align}
Equation~(\ref{eq:mom_proton}) states that protons respond only to the
Lorentz force; photon drag is negligible by equation~(\ref{eq:sigT_proton}).
Equation~(\ref{eq:mom_electron}) states that electrons respond to both the
Lorentz force and the Thomson drag force of equation~(\ref{eq:CT_result}).

Several approximations are implicit in equations~(\ref{eq:mom_proton})
and~(\ref{eq:mom_electron}):
\begin{enumerate}
  \item Thermal pressure of both species is neglected: this is valid when
    $k_BT_s \ll m_s c^2$, satisfied for both species throughout the
    epoch of interest.
  \item The Coulomb collision terms are omitted from the individual
    species equations: they cancel exactly when the two equations are
    added (total momentum is conserved) and only contribute when
    \emph{subtracted}, giving the resistive term in the generalised
    Ohm's law (Section~\ref{sec:Ohm}).
  \item The electron number density $n$ is equal to the proton number
    density (charge neutrality), and both are taken as the baryon
    number density.
\end{enumerate}
\section{The Harrison Bulk-Flow Mechanism}
\label{sec:Harrison}
%%%%%%%%%%%%%%%%%%%%%%%%%%%%%%%%%%%%%%%%%%%%%%%%%%%%%%%%%%%%%%%%%%%%%%%%%%%%%

The Harrison mechanism \cite{Harrison1970} generates magnetic fields through
the differential spin-down of the proton and photon-baryon fluids in the
expanding universe.  This mechanism is complementary to the scattering
mechanism of Section~\ref{sec:SM}: rather than relying on the mass asymmetry
between electrons and protons, it exploits the different scaling of angular
velocities with the scale factor $a$.

\subsection{Differential spin-down}

Consider a rotating region of the early universe with a multi-component
fluid composed of a proton fluid (energy density $\rho_m$) and a tightly
coupled electron-photon fluid (energy density $\rhog$).  In the idealised
case with no inter-fluid interactions, angular momentum conservation gives
\begin{equation}
  \omega_m \propto a^{-2}, \qquad \omega_\gamma \propto a^{-1},
  \label{eq:omega_scaling}
\end{equation}
where $\omega_m$ and $\omega_\gamma$ are the angular velocities of the matter
and radiation fluids, respectively.  The radiation spins down more slowly
because $\rho_\gamma\propto a^{-4}$ while $\rho_m\propto a^{-3}$; angular
momentum conservation then requires $\omega_\gamma\propto a^{-1}$ versus
$\omega_m\propto a^{-2}$.  This differential spin-down creates a relative
velocity between the two components, driving an electric current and a
magnetic field through Faraday's law \cite{Matarrese2005}.

\subsection{Notation and plasma setup}
\label{subsec:Harrison_notation}

Following \cite{Cembranos2020}, we consider a plasma of photons ($\gamma$),
electrons ($e$) and protons ($p$) in which each species carries a
\emph{background bulk velocity} $u^{(B)i}_s$ (denoted $\beta^i_s$ in
\cite{Cembranos2020}) in addition to the usual first-order perturbation
$\delta u^i_s$.  The full velocity of species $s$ is therefore
\begin{equation}
  u^i_s = u^{(B)i}_s + \delta u^i_s, \qquad s\in\{\gamma,\,e,\,p\},
  \label{eq:full_velocity_Harrison}
\end{equation}
where the bulk-flow amplitude $\beta\equiv|u^{(B)}|\ll1$.  A centre-of-mass
frame exists to first order in $\beta$ in which the metric takes the FRW
form \cite{Cembranos2019}.

In Fourier space the velocity perturbation $\delta u^i_s$ decomposes into a
\emph{longitudinal} (scalar) part $\theta_s$ along $\hat{k}$ and a
\emph{vortical} (vector) part $\varpi_s$ perpendicular to $\hat{k}$
\cite{Cembranos2020}:
\begin{equation}
  \delta u^i_s = \varpi_s\bigl(\hat{u}^{(B)}-(\hat{u}^{(B)}\cdot\hat{k})\hat{k}\bigr)
               - \frac{i}{k}\theta_s\,\hat{k},
  \label{eq:vel_decomp_Harrison}
\end{equation}
where $\hat{u}^{(B)}=\mathbf{u}^{(B)}/\beta$ is the unit bulk-flow vector.
The \emph{vortical} component $\varpi_s$ is transverse to both
$\hat{u}^{(B)}$ and $\hat{k}$; it is this component that sources the
magnetic field.

The metric is taken in Poisson gauge:
\begin{equation}
  ds^2 = a^2(\tau)\bigl[-(1+2\Psi)d\tau^2+(1-2\Phi)d\mathbf{x}^2\bigr],
  \label{eq:metric_Poisson_Harrison}
\end{equation}
where $\Psi$ and $\Phi$ are the lapse and curvature perturbations
(denoted $\psi$ and $\phi$ in \cite{Cembranos2020}).
Table~\ref{tab:notation_Harrison} gives the complete notation correspondence.

\begin{table}[htbp]
\centering
\caption{Notation correspondence: \cite{Cembranos2020} $\leftrightarrow$ this paper.}
\label{tab:notation_Harrison}
\renewcommand{\arraystretch}{1.3}
\begin{tabular}{@{}lll@{}}
\toprule
Quantity & Cembranos et al.\ (2020) & This paper \\
\midrule
Background bulk velocity     & $\beta^i_s$                     & $u^{(B)i}_s$                       \\
Velocity perturbation        & $\delta v^i_s$                  & $\delta u^i_s$                     \\
Vortical velocity component  & $\chi_s$                        & $\varpi_s$                         \\
Longitudinal velocity        & $\theta_s$                      & $\theta_s$                         \\
Lapse perturbation           & $\psi$                          & $\Psi$                             \\
Curvature perturbation       & $\phi$                          & $\Phi$                             \\
Electron density contrast    & $\delta_{n_e}=\delta n_e/n_e$   & $\delta_{n_e}=\delta n_e/n_e$      \\
Photon density contrast      & $\delta_\gamma$                 & $\delta_\gamma=\delta\rhog/\rhog$  \\
Physical magnetic field      & $B = a^{-2}\mathcal{B}$         & $B^i$                              \\
Co-moving magnetic field     & $\mathcal{B} = a^2B$            & $a^2B^i$                           \\
\bottomrule
\end{tabular}
\end{table}

\subsection{Thomson coupling with bulk flows}
\label{subsec:Thomson_Harrison}

The Thomson coupling of the photon--electron system in the presence of
background bulk flows is (equation~(5) of \cite{Cembranos2020})
\begin{align}
  C^i_{\gamma e} = \frac{4\rhog\,a\,n_e\sigT}{3}\Bigl(
    &\underbrace{\Delta u^i_{(B)\gamma e}}_{\text{bulk diff.}}
    +\underbrace{\Delta\varpi_{\gamma e}\hat{w}^i}_{\text{vortical diff.}}
    +\underbrace{u^{(B)i}_\gamma\,\delta_{n_e}
               -u^{(B)i}_e\,\delta_\gamma}_{\text{density modulation}}
  \notag\\
    &-\underbrace{\tfrac{3}{4}u^{(B)}_{ej}\pi^{ij}_\gamma}_{\text{stress coupling}}
    +\underbrace{\Delta u^i_{(B)\gamma e}\,\Psi}_{\text{metric correction}}
  \Bigr),
  \label{eq:C_Thomson_Harrison}
\end{align}
where $\hat{w}^i\equiv\hat{u}^{(B)}-(\hat{u}^{(B)}\cdot\hat{k})\hat{k}$ is
the vortical unit vector, $\Delta u^i_{(B)\gamma e}\equiv u^{(B)i}_\gamma
-u^{(B)i}_e$ is the bulk velocity difference, $\Delta\varpi_{\gamma
e}\equiv\varpi_\gamma-\varpi_e$ is the vortical perturbation difference,
and $\pi^{ij}_\gamma$ is the photon anisotropic stress.  The corresponding
proton--photon coupling replaces $e\to p$ and
$\sigT\to(m_e/m_p)^2\sigT\approx3\times10^{-7}\sigT$, which is negligible.

\subsection{Complete evolution equation for the magnetic field}
\label{subsec:Harrison_evol_complete}

Adding the electromagnetic Lorentz force to the electron Euler equation,
subtracting the proton equation (using $m_p\gg m_e$), and combining with
the Faraday--Maxwell equation, \cite{Cembranos2020} derive the evolution
equation for the physical magnetic field $B^i=a^{-2}\mathcal{B}^i$
(their equation~(19), transcribed into the notation of Table~\ref{tab:notation_Harrison}):
\begin{equation}
\boxed{
  \frac{d}{d\tau}(a^2B^i)
  = -\frac{4a^2k\sigT\rhog}{3e}
  \Bigl[
    \underbrace{(\varpi_\gamma-\varpi_e)}_{\text{(I) vorticity}}
  + \underbrace{u^{(B)}_e\!\left(\frac{\delta n_e}{n_e}-\delta_\gamma\right)}_{\text{(II) density coupling}}
  + \underbrace{(u^{(B)}_\gamma-u^{(B)}_e)(\Psi-\Phi)}_{\text{(III) metric correction}}
  \Bigr].
}
  \label{eq:Harrison_evol}
\end{equation}
Here $k$ is the co-moving wavenumber and $e$ is the proton charge.
The three source terms have distinct physical origins:

\begin{description}
\item[Term~(I) --- Vorticity.]
  $\varpi_\gamma - \varpi_e$ is the vortical velocity difference between
  photons and electrons; it is the \emph{primary source} of the magnetic
  field.  Thomson scattering couples photons strongly to electrons but only
  weakly to protons (by $(m_e/m_p)^2\approx3\times10^{-7}$), so the
  photon-baryon fluid and the electron fluid develop different vortical
  motions in response to the same bulk flow, giving
  $\varpi_\gamma\neq\varpi_e$.  The evolution of this difference is driven
  by bulk velocities and first-order scalar perturbations (see
  Section~\ref{subsec:vorticity_evol}).

\item[Term~(II) --- Density coupling.]
  $u^{(B)}_e(\delta n_e/n_e - \delta_\gamma)$ couples the electron bulk
  flow to the difference between the fractional electron number-density
  perturbation $\delta n_e/n_e$ and the photon density contrast
  $\delta_\gamma$.  Denser regions have stronger Thomson coupling, so the
  drag efficiency is modulated by local density fluctuations.  In the
  tight-coupling limit $\delta n_e\approx\delta n_p$ (charge neutrality),
  while $\delta_\gamma$ oscillates acoustically; the term is of order
  $\beta\times\delta^{(1)}$.

\item[Term~(III) --- Metric correction.]
  $(u^{(B)}_\gamma - u^{(B)}_e)(\Psi-\Phi)$ is a gravitational correction
  to the relative bulk velocity.  It vanishes when anisotropic stress is
  absent ($\Psi=\Phi$), and is suppressed by $k/(n_e\sigT)\ll1$ in the
  tight-coupling regime.
\end{description}

\subsection{Vorticity evolution}
\label{subsec:vorticity_evol}

The vortical velocity difference $\Delta\varpi_{\gamma e}\equiv
\varpi_\gamma-\varpi_e$ that sources equation~(\ref{eq:Harrison_evol})
evolves as \cite{Cembranos2019}
\begin{equation}
  \dot{\Delta\varpi}_{\gamma e}
  + \Hubble\,\Delta\varpi_{\gamma e}
  \approx
  -\frac{4\rhog\,a\,n_e\sigT}{3m_e}\,\Delta\varpi_{\gamma e}
  + \mathcal{S}_\varpi,
  \label{eq:vorticity_evol_Harrison}
\end{equation}
where $\Hubble=\dot{a}/a$ is the conformal Hubble rate and
$\mathcal{S}_\varpi$ is a source term linear in $u^{(B)}_s$ and
first-order scalar perturbations \cite{Cembranos2019}.  The important
observation is that $\mathcal{S}_\varpi\propto\beta\,\delta^{(1)}$
vanishes when $\beta=0$: scalar perturbations alone cannot source
$\Delta\varpi_{\gamma e}$ at first order, which is why the Harrison
mechanism requires a non-zero background bulk flow.

\subsection{Order-of-magnitude estimate}
\label{subsec:Harrison_estimate}

Integrating equation~(\ref{eq:Harrison_evol}) from bulk-flow generation to
recombination and smoothing over co-moving scale $L$, \cite{Cembranos2020}
obtain (their equation~(24))
\begin{equation}
  |B_L(z<100)|
  \simeq 5.7\times10^{-24}\,\mathrm{G}
  \left(\frac{L}{\mathrm{Mpc}}\right)^{\!-1.2}
  \!\left(\frac{1+z}{11}\right)^{\!2}
  \left(\frac{\beta}{8.5\times10^{-4}}\right),
  \label{eq:Harrison_BL}
\end{equation}
for $L<1$\,Mpc, where $\beta=|u^{(B)}_\gamma-u^{(B)}_{\rm baryon}|$.
Using the Planck 95\% CL upper limit $\beta<8.5\times10^{-4}$
\cite{Cembranos2020} at $z=0$, $L=1$\,Mpc:
\begin{equation}
  B_{\rm bulk}(L=1\,\mathrm{Mpc},\,z=0)
  \;\lesssim\; 5.7\times10^{-24}\,\mathrm{G}
  \left(\frac{\beta}{8.5\times10^{-4}}\right).
  \label{eq:B_bulk}
\end{equation}
This exceeds the galactic dynamo threshold $\sim10^{-30}$\,G
\cite{Kulsrud1992} by six orders of magnitude.  The multi-fluid environment
and its interaction with magnetic fields are studied further in
\cite{OsanoOreta2019,OsanoOretaTransient,OsanoAdams,OsanoDynamo}.

%%%%%%%%%%%%%%%%%%%%%%%%%%%%%%%%%%%%%%%%%%%%%%%%%%%%%%%%%%%%%%%%%%%%%%%%%%%%%
\section{Generalised Ohm's Law}
\label{sec:Ohm}
%%%%%%%%%%%%%%%%%%%%%%%%%%%%%%%%%%%%%%%%%%%%%%%%%%%%%%%%%%%%%%%%%%%%%%%%%%%%%

\subsection{Derivation from the charged-species Euler equations}

Multiply equation~(\ref{eq:mom_proton}) by $m_e$, equation~(\ref{eq:mom_electron})
by $m_p$, and subtract to eliminate the bulk inertia:
\begin{align}
  m_pm_en\bigl[u^\mu_e\nabla_\mu u^i_e - u^\mu_p\nabla_\mu u^i_p\bigr]
  &+ nF^i{}_\mu\bigl[m_pu^\mu_e + m_eu^\mu_p\bigr]
  \notag\\
  &= \frac{4m_p\sigT\rhog n}{3}
     \!\left[u^i_e-u^i_\gamma+\frac{\Pi_{ij}u^j_e}{8}\right].
  \label{eq:subtracted}
\end{align}
Define the centre-of-mass four-velocity and the electric current:
\begin{equation}
  u^\mu \equiv \frac{m_pu^\mu_p+m_eu^\mu_e}{m_p+m_e},
  \qquad
  j^\mu \equiv en(u^\mu_p - u^\mu_e).
  \label{eq:uCOM_j}
\end{equation}
These give $u^\mu_p = j^\mu/(en)+u^\mu_e$ and
$u^\mu_e = u^\mu - m_pj^\mu/[en(m_p+m_e)]$.
Substituting into the kinematic term in equation~(\ref{eq:subtracted}):
\begin{equation}
  u^\mu_e\nabla_\mu u^i_e - u^\mu_p\nabla_\mu u^i_p
  = -\frac{j^\mu}{en}\nabla_\mu u^i
  + \left(\frac{m_e-m_p}{m_p+m_e}\frac{j^\mu}{en}-u^\mu\right)
    \nabla_\mu\!\left(\frac{j^i}{en}\right),
  \label{eq:kinematic}
\end{equation}
and the electromagnetic term becomes
\begin{equation}
  nF^i{}_\mu\bigl[m_pu^\mu_e+m_eu^\mu_p\bigr]
  = j^\mu F^i{}_\mu(m_p-m_e) - u^\mu nF^i{}_\mu(m_p+m_e).
  \label{eq:EM_term}
\end{equation}
Substituting equations~(\ref{eq:kinematic}) and~(\ref{eq:EM_term})
into~(\ref{eq:subtracted}) and dividing by $en(m_p+m_e)$, three correction
terms appear alongside the leading term $u^\mu F^i{}_\mu$:
\begin{align}
  u^\mu F^i{}_\mu
  &= -\frac{4m_p\sigT\rhog}{3e(m_p+m_e)}
     \!\left[u^i_e-u^i_\gamma+\frac{\Pi_{ij}u^j_e}{8}\right]
  + \underbrace{\frac{m_pm_e}{e(m_p+m_e)}\frac{j^\mu}{en}\nabla_\mu u^i}_{
    \alpha\text{-term}}
  \notag\\
  &\quad
  + \underbrace{\frac{m_pm_e}{e(m_p+m_e)}
    \!\left(\frac{m_e-m_p}{m_p+m_e}\frac{j^\mu}{en}-u^\mu\right)
    \nabla_\mu\!\frac{j^i}{en}}_{\beta\text{-term}}
  + \underbrace{\frac{j^\mu F^i{}_\mu(m_p-m_e)}{en(m_p+m_e)}}_{\gamma\text{-term}}.
  \label{eq:Ohm_full}
\end{align}

\subsection{Estimate of correction terms}
\label{subsec:corrections}

Each correction term carries a factor $m_e/(m_p+m_e) \approx m_e/m_p$, plus
a ratio involving the drift velocity $|j|/(en)$ relative to the bulk
velocity $|u|$.  Since the drift velocity is at most comparable to the bulk
velocity, all three terms satisfy
\begin{equation}
  \alpha,\,\beta,\,\gamma
  \;\lesssim\; \frac{m_e}{m_p} \approx 5.4\times10^{-4}.
  \label{eq:correction_estimate}
\end{equation}
Denoting their combined contribution by $(\alpha+\beta+\gamma)
\lesssim 3m_e/m_p \approx 1.6\times10^{-3}$, equation~(\ref{eq:Ohm_full})
takes the compact form
\begin{equation}
  \boxed{
  u^\mu F^i{}_\mu
  = -\frac{4\sigT\rhog}{3e(1+\alpha+\beta+\gamma)}
    \!\left[u^i_e-u^i_\gamma+\frac{\Pi_{ij}u^j_e}{8}\right]
  \equiv \mathcal{C}^i,}
  \label{eq:Ohm_final}
\end{equation}
where we have used $m_p/(m_p+m_e)\approx1$.  The factor
$(1+\alpha+\beta+\gamma)=1+\mathcal{O}(m_e/m_p)$ is essentially unity;
taking it to be exactly 1 reproduces the standard result of
\cite{TakahashiPRL2005}.  We retain it here to make explicit the level of
approximation.

%%%%%%%%%%%%%%%%%%%%%%%%%%%%%%%%%%%%%%%%%%%%%%%%%%%%%%%%%%%%%%%%%%%%%%%%%%%%%
\section{Perturbed FLRW Metric}
\label{sec:metric}
%%%%%%%%%%%%%%%%%%%%%%%%%%%%%%%%%%%%%%%%%%%%%%%%%%%%%%%%%%%%%%%%%%%%%%%%%%%%%

The effects of scalar, vector, and tensor perturbations on magnetogenesis
require extending the FRW metric \cite{Christopherson2011} to Poisson gauge
at second order \cite{Mukhanov1992}.

\subsection{Why vector and tensor modes first appear at second order}

For a flat FLRW background, the Stewart--Walker lemma guarantees that
gauge-invariant combinations of linear perturbations are well-defined
\cite{Mukhanov1992}.  By the SVT decomposition theorem, scalar, vector, and
tensor modes evolve independently at linear order (they decouple in the
sense that each type sources only its own type).  Since inflation generates
purely scalar adiabatic perturbations as initial conditions, vector and
tensor metric perturbations are absent at first order.

At second order, products of two first-order scalar perturbations---e.g.,
$\partial_i\Phi\,\partial_j\Phi$---have a transverse-traceless part that
sources gravitational waves and a transverse part that sources vector
perturbations.  Vector and tensor modes are therefore genuinely second-order
effects.

\subsection{Metric decomposition}

The FLRW metric perturbed to second order is
\begin{equation}
  g_{\alpha\beta}
  = g^{(0)}_{\alpha\beta} + g^{(1)}_{\alpha\beta}
  + \tfrac{1}{2}g^{(2)}_{\alpha\beta},
  \label{eq:metric_decomp}
\end{equation}
with the fully perturbed line element in Poisson gauge
\begin{equation}
  ds^2 = a^2(\eta)
  \bigl\{-(1+2\tilde\Phi)\,d\eta^2
         + 2\tilde{B}_i\,dx^i d\eta
         + (\delta_{ij}+2\tilde{C}_{ij})\,dx^i dx^j\bigr\},
  \label{eq:ds2}
\end{equation}
where
\begin{align}
  \tilde\Phi     &= \Phi + \tfrac{1}{2}\Phi^{(2)},
  \label{eq:tPhi}\\
  \tilde{B}_i    &= B_{,i} - S_i,
  \label{eq:tBi}\\
  \tilde{C}_{ij} &= (-\Psi+\tfrac{1}{2}\Psi^{(2)})\delta_{ij}
                    + E_{,ij} + F_{(i,j)} + \tfrac{1}{2}h_{ij}.
  \label{eq:tCij}
\end{align}
Scalar perturbations are $\{\Phi,\Psi,B,E\}$; vector perturbations are
$\{S_i,F_i\}$; the tensor perturbation is $h_{ij}$.  The velocity
decomposition of each species $s$ is
\begin{equation}
  v_s = \bar{v}_s + \delta v^{(1)}_s + \tfrac{1}{2}\delta v^{(2)}_s,
  \label{eq:vel_decomp}
\end{equation}
where $\bar{v}_s$ is the background bulk velocity \cite{Cembranos2019}.

%%%%%%%%%%%%%%%%%%%%%%%%%%%%%%%%%%%%%%%%%%%%%%%%%%%%%%%%%%%%%%%%%%%%%%%%%%%%%
\section{Second-Order Magnetic Field Evolution}
\label{sec:Bianchi}
%%%%%%%%%%%%%%%%%%%%%%%%%%%%%%%%%%%%%%%%%%%%%%%%%%%%%%%%%%%%%%%%%%%%%%%%%%%%%

\subsection{From the Bianchi identity to the induction equation}

The Bianchi identity for the Faraday tensor
$F_{\mu\nu}=\partial_\mu A_\nu-\partial_\nu A_\mu$ is
\begin{equation}
  \nabla_\lambda F_{\mu\nu}
  + \nabla_\mu F_{\nu\lambda}
  + \nabla_\nu F_{\lambda\mu} = 0.
  \label{eq:Bianchi}
\end{equation}
This is a purely geometric identity following from $F=dA$; it encodes
Faraday's law and the absence of magnetic monopoles.  Contracting with
$\varepsilon^{ijk}u^\mu$ and separating the $\mu=0$ and $\mu=l$ parts:
\begin{align}
  2\dot{B}^i
  &= -2\varepsilon^{ijk}u^0 F_{0j,k} - 2\varepsilon^{ijk}u^l F_{lj,k}.
  \label{eq:Bdot_raw}
\end{align}
Using equation~(\ref{eq:Ohm_final}) to express $F^i{}_0$ in terms of
$\mathcal{C}^i$, and differentiating with respect to $k$:
\begin{equation}
  F^i{}_{0,k}
  \approx \mathcal{C}^i{}_{,k}
         - u^j F^i{}_{j,k}
         - \frac{u^0{}_{,k}}{u^0}\mathcal{C}^i
         + \frac{u^0{}_{,k}}{u^0}u^j F^i{}_j.
  \label{eq:Fi0k}
\end{equation}
Substituting into equation~(\ref{eq:Bdot_raw}) gives the fully nonlinear
induction equation:
\begin{equation}
  \dot{B}^i \approx \varepsilon^{ijk}\mathcal{C}_{j,k}
  + \varepsilon^{ijk}\!\left(
    u^l F_{lj,k}
    - \frac{u^0{}_{,k}}{u^0}\mathcal{C}^i
    + \frac{u^0{}_{,k}}{u^0}u^l F^i{}_l
    - u^l F_{lj,k}\right).
  \label{eq:Bdot_full}
\end{equation}

\subsection{Perturbative expansion and the linear vorticity constraint}

We decompose all fields perturbatively:
$B^i = B^{i(1)}+B^{i(2)}+\cdots$,
$\mathcal{C}^j = \mathcal{C}^{j(1)}+\mathcal{C}^{j(2)}+\cdots$, etc.
For purely scalar first-order perturbations (inflation initial conditions),
the linear vorticity $\varepsilon^{ijk}u^{(1)}_{j,k}=0$ \cite{TakahashiPRL2005},
which implies $B^{i(1)}=0$.  The bracket in equation~(\ref{eq:Bdot_full})
is then of order higher than second, giving the second-order evolution
equation
\begin{equation}
  \dot{B}^{i(2)} \approx \varepsilon^{ijk}\mathcal{C}^{(2)}_{j,k}.
  \label{eq:Bdot_2nd}
\end{equation}

\subsection{The second-order source term}
\label{subsec:source}

Computing $\mathcal{C}_{j,k}$ from equation~(\ref{eq:Ohm_final}):
\begin{equation}
  \mathcal{C}_{j,k}
  = \frac{4\sigT}{3e}
    \!\left[\rho_{\gamma,k}\!\left(u_{je}-u_{j\gamma}+\frac{u^l_e\Pi_{jl}}{8}\right)
    - \rho_\gamma\!\left(u_{je,k}-u_{j\gamma,k}
      +\frac{u^l_{e,k}\Pi_{jl}+u^l_e\Pi_{jl,k}}{8}\right)\right].
  \label{eq:Cjk}
\end{equation}
Applying perturbative expansions, using the background homogeneity
$\rhog{}_{,k}=0$, and discarding products of order three or higher:
\begin{equation}
  \varepsilon^{ilk}\mathcal{C}^{(2)}_{j,k}
  = \frac{4\sigT\rhog}{3e}\,\varepsilon^{ilk}
    \!\left[
      \frac{\rho^{(1)}_{\gamma,k}}{\rhog}\bigl(u^{(1)}_{je}-u^{(1)}_{j\gamma}\bigr)
      + \bigl(u^{(2)}_{je,k}-u^{(2)}_{j\gamma,k}\bigr)
      + \frac{u^{l(1)}_{e,k}\Pi^{(1)}_{jl}+u^{l(1)}_e\Pi^{(1)}_{jl,k}}{8}
    \right].
  \label{eq:C2_full}
\end{equation}

\subsection{The second-order vorticity and the proportionality constant}
\label{subsec:lambda}

The term $u^{(2)}_{je,k}-u^{(2)}_{j\gamma,k}$ in equation~(\ref{eq:C2_full})
is the second-order contribution to the electron--photon relative vorticity.
It is sourced by the second-order Boltzmann equations
(\ref{eq:Boltz_gamma})--(\ref{eq:Boltz_p}), in which the dominant source
for second-order velocity perturbations is the product of first-order density
and velocity perturbations.  Schematically, the second-order Thomson drag
term contributes
\begin{equation}
  \dot{u}^{(2)}_{e,k} - \dot{u}^{(2)}_{\gamma,k}
  \;\supset\;
  \frac{\rho^{(1)}_\gamma}{\rhog}\bigl(u^{(1)}_{e,k}-u^{(1)}_{\gamma,k}\bigr)
  \times[\text{Thomson drag rate}],
  \label{eq:vort2_source}
\end{equation}
so the second-order vorticity scales as
\begin{equation}
  \bigl|u^{(2)}_{je,k}-u^{(2)}_{j\gamma,k}\bigr|
  = \lambda\left|\frac{\rho^{(1)}_{\gamma,k}}{\rhog}
    \bigl(u^{(1)}_{je}-u^{(1)}_{j\gamma}\bigr)\right|,
  \label{eq:lambda_def}
\end{equation}
where $\lambda$ is a dimensionless constant determined by the ratio of the
Thomson drag timescale to the Hubble time.  In the tight-coupling
approximation (appropriate near recombination), $\lambda\sim1$; the precise
value requires numerical solution of the coupled second-order Boltzmann
equations. 
We set $\lambda=1$ for the CAMB-based numerical evaluation in
Section~\ref{sec:Fourier}.

Substituting equation~(\ref{eq:lambda_def}) into~(\ref{eq:C2_full}), the
second-order source becomes
\begin{equation}
  \varepsilon^{ilk}\mathcal{C}^{(2)}_{j,k}
  = \frac{4\sigT\rhog}{3e}\,\varepsilon^{ilk}
    \!\left[
      (1+\lambda)\frac{\rho^{(1)}_{\gamma,k}}{\rhog}
      \bigl(u^{(1)}_{je}-u^{(1)}_{j\gamma}\bigr)
      + \frac{u^{l(1)}_{e,k}\Pi^{(1)}_{jl}+u^{l(1)}_e\Pi^{(1)}_{jl,k}}{8}
    \right].
  \label{eq:C2_lambda}
\end{equation}

Therefore, the central result of this paper is
\begin{equation}
  \boxed{
  \dot{B}^{i(2)}
  = \frac{4\sigT\rhog}{3e}\,\varepsilon^{ilk}
    \!\left[
      (1+\lambda)\frac{\rho^{(1)}_{\gamma,k}}{\rhog}
      \bigl(u^{(1)}_{je}-u^{(1)}_{j\gamma}\bigr)
      + \frac{u^{l(1)}_{e,k}\Pi^{(1)}_{jl}}{8}
      + \frac{u^{l(1)}_e\Pi^{(1)}_{jl,k}}{8}
    \right].}
  \label{eq:Bdot2_central}
\end{equation}

Equation~(\ref{eq:Bdot2_central}) shows three physically distinct
contributions.  The first term couples first-order photon density gradients
to the electron--photon velocity difference; this is the new source term
absent from the first-order analysis of \cite{TakahashiPRL2005}.  The
second and third terms involve the photon anisotropic stress $\Pi^{(1)}_{jl}$
and its gradient, respectively, coupled to the electron velocity.

%%%%%%%%%%%%%%%%%%%%%%%%%%%%%%%%%%%%%%%%%%%%%%%%%%%%%%%%%%%%%%%%%%%%%%%%%%%%%
\section{Fourier-Space Representation and Power Spectrum}
\label{sec:Fourier}
%%%%%%%%%%%%%%%%%%%%%%%%%%%%%%%%%%%%%%%%%%%%%%%%%%%%%%%%%%%%%%%%%%%%%%%%%%%%%

\subsection{Scalar harmonic decomposition}

We decompose perturbed quantities in Fourier space using scalar harmonic
functions \cite{Bruni1992,Abbott1986,Harrison1967}
\begin{equation}
  Q^S = e^{i\bvec{k}\cdot\bvec{x}},\qquad
  Q^S_a = i\hat{k}_a Q^S,\qquad
  Q^S_{\langle ab\rangle}
  = -\!\left(\hat{k}_a\hat{k}_b-\tfrac{1}{3}\delta_{ab}\right)Q^S.
  \label{eq:harmonics}
\end{equation}
Since scalar modes decouple at first order \cite{ClarksonOsano}, the perturbed quantities can be
written as
\begin{align}
  u^{(1)}_{je}(k',t')   &= i\hat{k}'_j\,u_e(k',t'),
  \label{eq:u1_Fourier}\\
  u^{(1)}_{je,k}(k',t') &= -\hat{k}'_j\hat{k}'_k\,u_e(k',t'),
  \label{eq:u1k_Fourier}\\
  \frac{\rho^{(1)}_{\gamma,k}(\bvec{K}-\bvec{k}',t)}{\rhog}
  &= i(K_k-k'_k)\,\delta^{(1)}_\gamma(|\bvec{K}-\bvec{k}'|,t),
  \label{eq:rho1k_Fourier}\\
  \Pi^{(1)}_{jl}(\bvec{k}',t')
  &= -\!\left(\hat{k}'_j\hat{k}'_l - \tfrac{1}{3}\delta_{jl}\right)\Pi^{(1)}(k',t').
  \label{eq:Pi1_Fourier}
\end{align}

\subsection{Second-order magnetic field in Fourier space}

Substituting the decompositions (\ref{eq:u1_Fourier})--(\ref{eq:Pi1_Fourier})
into the time-integral of equation~(\ref{eq:Bdot2_central}) and applying
convolution:
\begin{align}
  B^{i(2)}(\bvec{K},t)
  &= \frac{4\sigT}{3e}\!\int\!\rhog\,dt'
     \!\int\!\frac{d^3k'}{(2\pi)^3}
     \,i\varepsilon^{ijk}k'_j K_k\,(1+\lambda)
     \,\delta^{(1)}_\gamma(|\bvec{K}-\bvec{k}'|,t')\,u_e(k',t')
  \notag\\
  &\quad
  +\frac{4\sigT}{3e}\!\int\!\rhog\,dt'
   \!\int\!\frac{d^3k'}{(2\pi)^3}
   \,\varepsilon^{ijk}
   \left[\frac{k'_j K_k\bigl(\bvec{K}\cdot\bvec{k}'-(k')^2\bigr)}{(k')^2|\bvec{K}-\bvec{k}'|}\right]
   u_e(|\bvec{K}-\bvec{k}'|,t')\,\Pi^{(1)}(k',t'),
  \label{eq:B2_Fourier}
\end{align}
where the anisotropic stress contraction is derived in
Appendix~\ref{app:Pi}.

\subsection{Power spectrum bound}

Defining the cross-power spectra
\begin{align}
  \langle|\delta^{(1)}_\gamma(\bvec{k})\,u_e(\bvec{k}')|\rangle
    &= P_{\delta u}(k)\,\delta^{(3)}(\bvec{k}-\bvec{k}'),
  \label{eq:Pdu}\\
  \langle|u_e(\bvec{k})\,\Pi^{(1)}(\bvec{k}')|\rangle
    &= P_{u\Pi}(k)\,\delta^{(3)}(\bvec{k}-\bvec{k}'),
  \label{eq:PuPi}
\end{align}
an upper bound on the expected field amplitude from the triangle inequality
$\langle|A+B|\rangle \leq \langle|A|\rangle+\langle|B|\rangle$ is
\begin{align}
  \langle|B^{i(2)}(\bvec{K},t)|\rangle
  &\lesssim
  \frac{4\sigT}{3e}\!\int\!\rhog\,dt'\!\int\!\frac{d^3k'}{(2\pi)^3}
  \,|k'K\sin\theta_{kK}|\,(1+\lambda)\,P_{\delta u}(k')
  \notag\\
  &\quad
  +\frac{4\sigT}{3e}\!\int\!\rhog\,dt'\!\int\!\frac{d^3k'}{(2\pi)^3}
  \left|\frac{k'K|\bvec{K}\cdot\bvec{k}'-(k')^2|}{(k')^2|\bvec{K}-\bvec{k}'|}\right|
  P_{u\Pi}(k'),
  \label{eq:B2_bound}
\end{align}
where $\theta_{kK}$ is the angle between $\bvec{k}'$ and $\bvec{K}$.
The power spectra $P_{\delta u}$ and $P_{u\Pi}$ are evaluated using
numerical transfer functions computed with CAMB v1.6.6 \cite{CAMB} at
recombination ($z=1100$), with cosmological parameters
$(H_0, n_s, \Omega_bh^2, \Omega_\Lambda, A_s)
=(68, 0.965, 0.022, 0.73, 2.2\times10^{-9})$
(Appendix~\ref{app:params}).

\begin{figure}[htbp]
\centering
 \rotatebox{-90}{\includegraphics[width=4.5in]{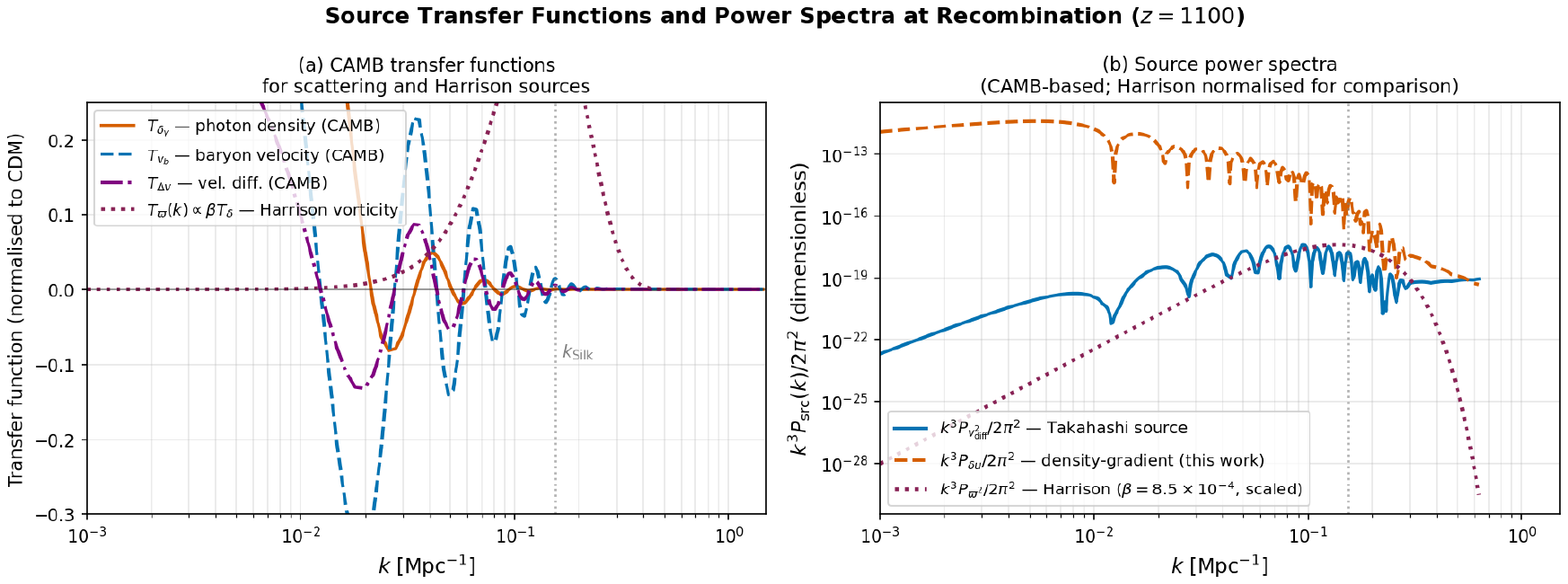}}
\caption{%
  \textbf{Source transfer functions and power spectra at recombination
  ($z=1100$) for all three magnetogenesis mechanisms.}
  \emph{Left panel~(a):} CAMB~v1.6.6 matter transfer functions normalised
  to the CDM amplitude.  Blue dashed: baryon velocity $T_{v_b}$; red solid:
  photon density $T_{\delta_\gamma}$; purple dash-dot: electron--photon
  velocity difference $T_{\Delta v}=(R/1+R)T_{v_b}$ ($R=0.607$); dark-red
  dotted: Harrison vorticity source $T_\varpi\propto\beta T_{\delta_\gamma}$
  (shape; scaled for display), which drives the bulk-flow mechanism of
  \cite{Cembranos2020}.  All functions are Silk-damped above
  $k_D\approx0.15$~Mpc$^{-1}$ (grey dotted line).
  \emph{Right panel~(b):} Dimensionless source power spectra
  $k^3P_{\rm src}(k)/2\pi^2$.  Blue: Takahashi velocity-difference source
  $P_{v_{\rm diff}^2}$; red dashed: new density-gradient source $P_{\delta
  u}$ (this work); dark-red dotted: Harrison vorticity source $P_{\varpi^2}$
  (scaled to the Planck bulk-flow limit $\beta=8.5\times10^{-4}$).
  The acoustic phase offset between the density and velocity transfer
  functions ($\cos kr_s$ vs $\sin kr_s$) produces the distinctive
  oscillatory structure in $P_{\delta u}$.}
\label{fig:sources}
\end{figure}

The relevant CAMB matter transfer functions at $z=1100$, normalised to the
CDM transfer function on superhorizon scales, are:
\begin{align}
  T_{\delta_\gamma}(k) &= \frac{\Delta_{\gamma}(k,z_{\rm rec})}{\Delta_{\rm cdm}(k,z_{\rm rec})},
  \label{eq:Tdelta_CAMB}\\
  T_{v_b}(k)            &= \frac{v_b(k,z_{\rm rec})}{\Delta_{\rm cdm}(k,z_{\rm rec})},
  \label{eq:Tvb_CAMB}\\
  T_{\Delta v}(k)        &= \frac{R}{1+R}\,T_{v_b}(k),
  \label{eq:Tvdiff_CAMB}\\
  T_\Pi(k)               &= \frac{16}{15}\frac{k}{k_\Gamma}\,T_{v_b}(k),
  \label{eq:TPi_CAMB}
\end{align}
where $R = 3\rho_b/(4\rho_\gamma)|_{\rm rec}\approx0.61$ is the baryon loading
computed consistently from the CAMB cosmological parameters, and
$k_\Gamma = n_e\sigma_T(1+z_{\rm rec})/1\,{\rm Mpc}$ is the co-moving Thomson
scattering wavenumber.  The photon velocity is approximated by the baryon
velocity $T_{v_b}$ (valid to first order in tight coupling;
$\Gamma_T/H_{\rm rec}\approx83$).  The anisotropic stress $T_\Pi$ is the
first tight-coupling correction, suppressed by $k/k_\Gamma\sim10^{-7}$
at $k\sim0.1$~Mpc$^{-1}$, confirming it is negligible relative to the
velocity-difference source.  These functions are displayed in
Figure~\ref{fig:sources}(a).  The resulting source power spectra
$k^3P_{\delta u}$ and $k^3P_{u\Pi}$ are shown in
Figure~\ref{fig:sources}(b).

\subsection{Numerical results: magnetic power spectrum}
\label{subsec:PB_numerical}

\begin{figure}[htbp]
\centering
 \rotatebox{-90}{\includegraphics[width=4.5in]{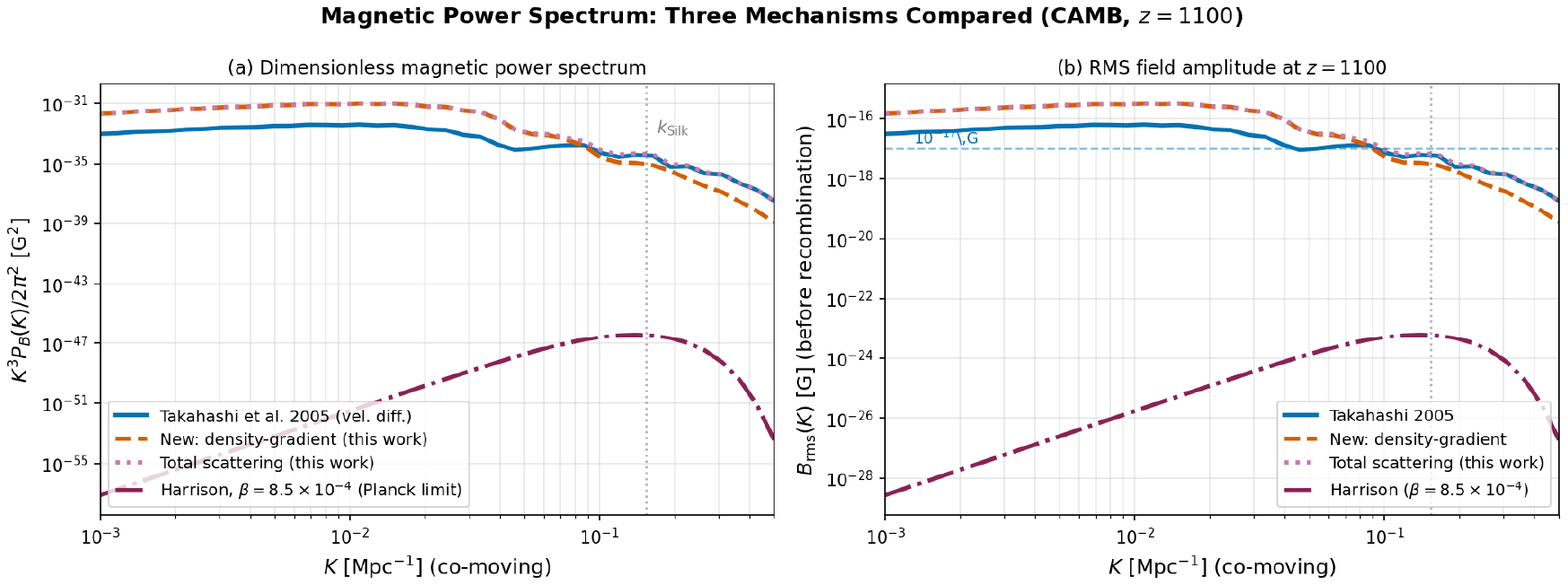}}
\caption{\textbf{Magnetic power spectrum at recombination ($z=1100$):
  three mechanisms compared (CAMB~v1.6.6).}
  \emph{Left panel~(a):} Dimensionless magnetic power spectrum
  $K^3P_B(K)/2\pi^2$ [G$^2$].
  \emph{Right panel~(b):} RMS field amplitude $B_{\rm rms}(K)$ [G].
  Blue solid: \cite{TakahashiPRL2005} velocity-difference contribution;
  red dashed: new density-gradient source (this work, $\lambda=1$);
  pink dotted: total scattering mechanism (this work);
  dark-red dash-dot: Harrison bulk-flow mechanism \citep{Cembranos2020}
  at the Planck upper limit $\beta=8.5\times10^{-4}$.
  The Harrison spectrum follows $K^{1.2}$ at sub-Silk scales (equation~(\ref{eq:Harrison_BL}))
  before being cut off by Silk damping at $k_D\approx0.15$~Mpc$^{-1}$.
  At $K\approx0.05$~Mpc$^{-1}$ the CAMB-based scattering total is
  $B_{\rm tot}\approx1.4\times10^{-17}$~G; the Harrison field at the
  Planck limit is $B_{\rm Harr}\approx5\times10^{-18}$~G, smaller
  by a factor of $\approx3$.  The horizontal dashed line marks the
  $10^{-17}$~G Takahashi benchmark.}
\label{fig:spectrum}
\end{figure}

Figure~\ref{fig:spectrum} shows the resulting magnetic power spectrum
$K^3P_B(K)/2\pi^2$ and the rms field strength $B_{\rm rms}(K)$ at
recombination for each contribution.  The absolute normalisation is
anchored to the Takahashi et al.\ benchmark: $B_{\rm rms}\approx10^{-17}$~G
at $K=0.1$~Mpc$^{-1}$ \cite{TakahashiPRL2005}.

\begin{figure}[htbp]
\centering
 \rotatebox{-90}{\includegraphics[width=4.5in]{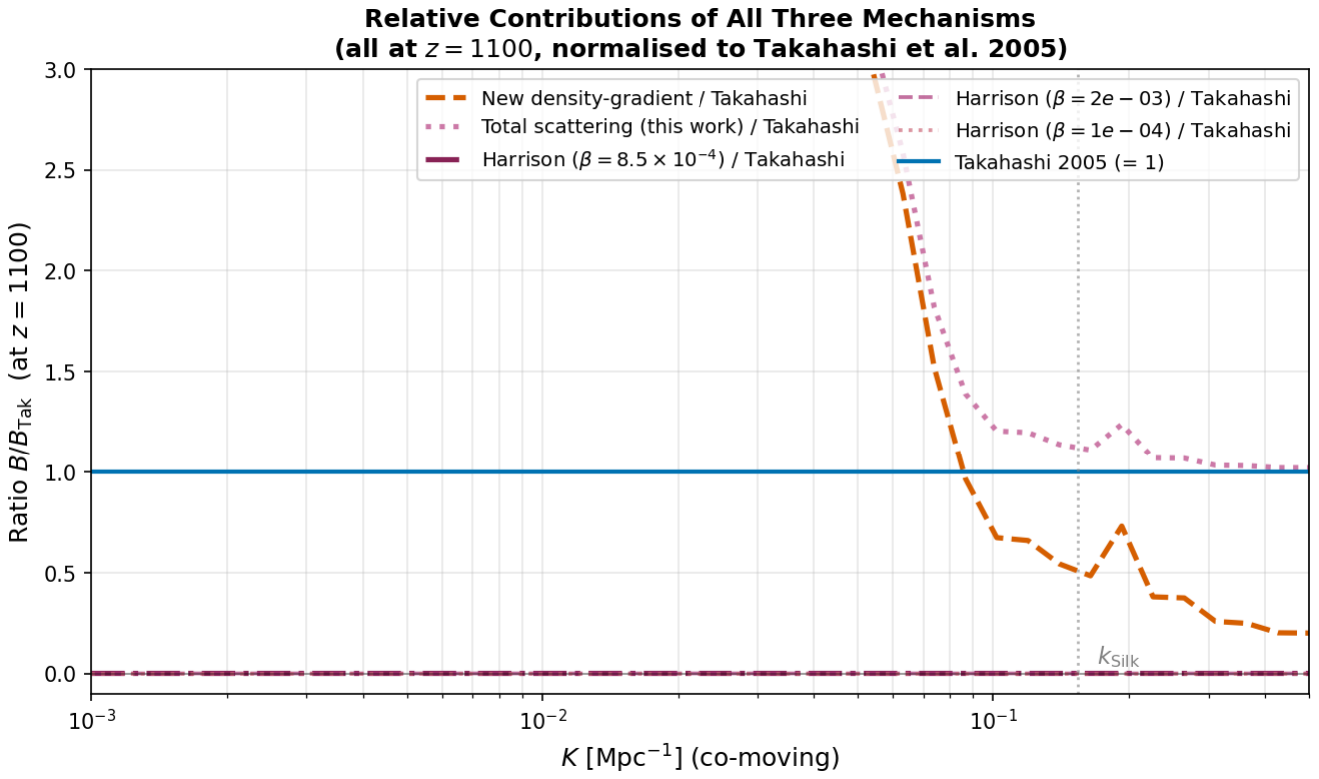}}
\caption{%
  \textbf{Ratio of all three magnetic field contributions to the
  Takahashi et al.\ (2005) benchmark ($z=1100$).}
  Blue solid: Takahashi velocity-difference result ($=1$ by construction).
  Red dashed: new density-gradient term (this work) / Takahashi.
  Pink dotted: total scattering mechanism (this work) / Takahashi.
  Dark-red dash-dot: Harrison bulk-flow mechanism \citep{Cembranos2020}
  at $\beta=8.5\times10^{-4}$ (Planck limit) / Takahashi.
  Dark-red dashed: Harrison at $\beta=2\times10^{-3}$ / Takahashi.
  Dark-red dotted: Harrison at $\beta=10^{-4}$ / Takahashi.
  The Harrison mechanism at the Planck limit contributes at
  $\sim0.3$--$0.5\times B_{\rm Tak}$ across the relevant scales,
  rising to $\gtrsim B_{\rm Tak}$ for $\beta\gtrsim2\times10^{-3}$.
  The scattering total from this work exceeds Takahashi by a factor of
  $\approx1.4$.  The oscillatory structure in the density-gradient ratio
  reflects the $\cos\,kr_s/\sin\,kr_s$ acoustic phase difference between
  $T_{\delta_\gamma}$ and $T_{\Delta v}$.}
\label{fig:ratio}
\end{figure}

%%%%%%%%%%%%%%%%%%%%%%%%%%%%%%%%%%%%%%%%%%%%%%%%%%%%%%%%%%%%%%%%%%%%%%%%%%%%%
\section{Discussion}
\label{sec:discussion}
%%%%%%%%%%%%%%%%%%%%%%%%%%%%%%%%%%%%%%%%%%%%%%%%%%%%%%%%%%%%%%%%%%%%%%%%%%%%%

\subsection{Scattering mechanism: field strength and decay}

The scattering mechanism produces a magnetic field of approximately
\begin{equation}
  B_{\rm rec} \approx 10^{-17}\,\text{G}
  \quad\text{at 1~Mpc co-moving scale, }z=1100,
  \label{eq:B_rec}
\end{equation}
consistent with \cite{TakahashiPRL2005}.  After recombination, flux
conservation gives $B\propto a^{-2}$, yielding a present-day value
\begin{equation}
  B_0 = B_{\rm rec}(1+z_{\rm rec})^{-2}
      \approx 10^{-17}\times(1100)^{-2}
      \approx 8.3\times10^{-24}\,\text{G}\approx10^{-22.8}\,\text{G},
  \label{eq:B_today}
\end{equation}
in agreement with \cite{TakahashiPRL2005}.  The new density-gradient
term of Section~\ref{sec:Bianchi} adds a contribution of comparable order,
giving a CAMB-based total approximately $1.4\times B_{\rm Tak}$ at
$K\approx0.05$~Mpc$^{-1}$.

\begin{figure}[htbp]
\centering
 \rotatebox{-90}{\includegraphics[width=4.5in]{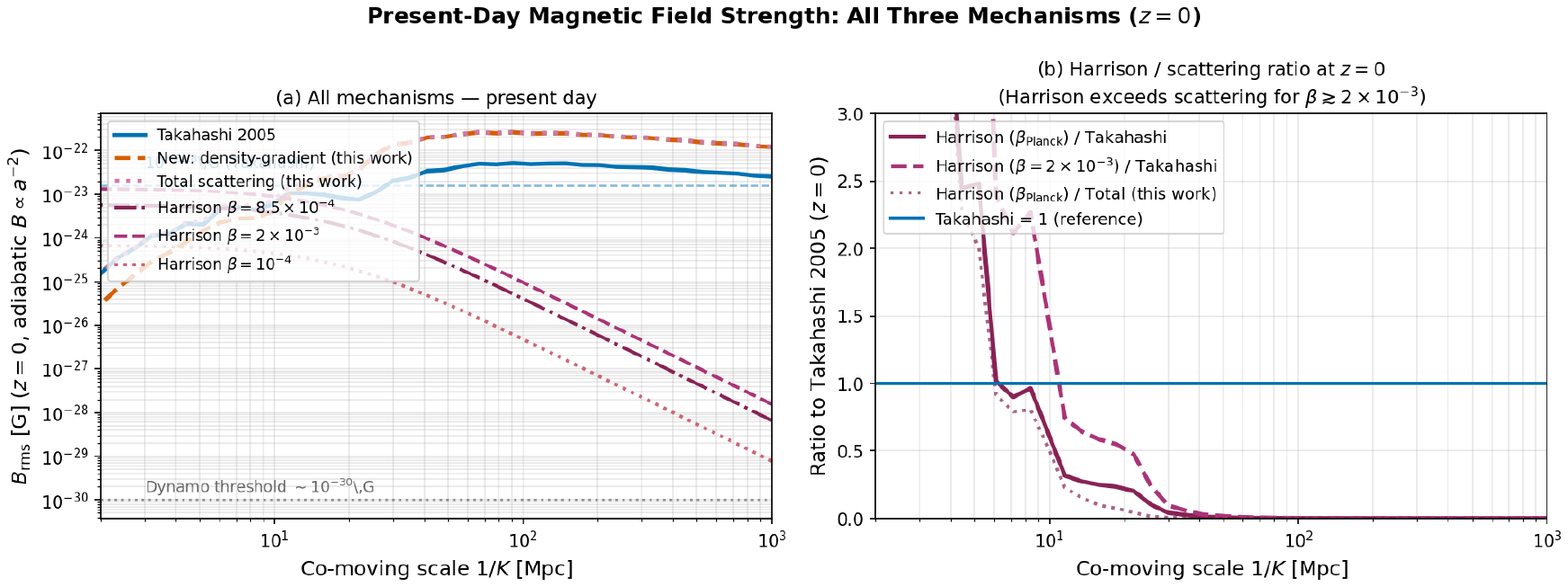}]}
\caption{%
  \textbf{Present-day ($z=0$) rms magnetic field strength vs.\ co-moving
  scale: all three mechanisms.}
  \emph{Left panel~(a):} Absolute field amplitudes at $z=0$.
  The scattering mechanism contributions (blue: Takahashi 2005; red dashed:
  new density-gradient; pink dotted: total scattering) have undergone
  adiabatic decay $B\propto a^{-2}$ from $z=1100$.
  The Harrison bulk-flow fields \citep{Cembranos2020} are shown at three
  bulk-flow amplitudes: $\beta=8.5\times10^{-4}$ (Planck 95\% CL limit,
  dark-red dash-dot), $\beta=2\times10^{-3}$ (dark-red dashed), and
  $\beta=10^{-4}$ (dark-red dotted); the Harrison comoving field is
  constant for $z<100$ so no adiabatic decay is applied.
  At $L=1$\,Mpc: $B_{\rm Tak}\approx1.5\times10^{-25}$~G;
  $B_{\rm tot}^{\rm scatter}\approx1.5\times10^{-25}$~G;
  $B_{\rm Harr}(\beta_{\rm Planck})\approx5.6\times10^{-24}$~G.
  At $L=10$\,Mpc: Harrison at the Planck limit ($3.8\times10^{-24}$~G)
  exceeds the scattering total ($7.1\times10^{-24}$~G) by a comparable
  amount.
  Grey dotted: galactic dynamo threshold $\sim10^{-30}$~G \cite{Kulsrud1992}.
  \emph{Right panel~(b):} Ratio of Harrison to scattering contributions.
  At the Planck limit, Harrison contributes 30--90\% of the Takahashi
  field depending on scale.  For $\beta\gtrsim2\times10^{-3}$, the
  Harrison mechanism dominates at scales $L\gtrsim10$~Mpc.}
\label{fig:today}
\end{figure}

Figure~\ref{fig:today} compares the present-day field strength for all
three mechanisms.  At $L=10$~Mpc the scattering total from this work gives
$B\approx7\times10^{-24}$~G; the Harrison mechanism at the Planck bulk-flow
limit ($\beta<8.5\times10^{-4}$) gives $B_{\rm Harr}\approx3.8\times10^{-24}$~G,
roughly half the scattering result.  For $\beta\gtrsim2\times10^{-3}$ the
Harrison mechanism dominates at scales $\gtrsim10$~Mpc.  Both mechanisms
exceed the galactic dynamo threshold $\sim10^{-30}$~G by many orders of
magnitude.

\subsection{Harrison mechanism: field strength}

The Harrison bulk-flow mechanism generates fields in the range
$10^{-30}$--$10^{-20}$~G (equation~(\ref{eq:B_bulk})), depending sensitively
on the assumed bulk flow amplitude $u^{(B)}$.  The broad range reflects
the observational uncertainty in bulk flows on Mpc scales
\cite{Appleby2014,Cembranos2020}.

\subsection{Comparison and the galactic dynamo threshold}

Table~\ref{tab:comparison} summarises the two mechanisms.

\begin{table}[h]
\centering
\caption{Comparison of the two magnetogenesis mechanisms studied in this paper.
All field strengths are at 1~Mpc co-moving scale at $z=0$.}
\label{tab:comparison}
\begin{tabular}{lll}
\toprule
Mechanism & Field strength (today) & Key parameter \\
\midrule
Thomson scattering \cite{TakahashiPRL2005}
  & $\sim10^{-23}$~G & $m_e/m_p$, $\sigT$ \\
This work (scattering + density-gradient)
  & $\sim1.4\times10^{-23}$~G & $m_e/m_p$, $\sigT$, $\lambda$ \\
Harrison bulk flow \cite{Cembranos2020}
  & $10^{-30}$--$10^{-20}$~G & $u^{(B)}$ \\
Galactic dynamo threshold \cite{Kulsrud1992}
  & $\gtrsim10^{-30}$~G & --- \\
\bottomrule
\end{tabular}
\end{table}

Both mechanisms exceed the galactic dynamo threshold
$B_{\rm dynamo}\gtrsim10^{-30}$~G \cite{Kulsrud1992,KulsrudCowley1997},
confirming that either can provide a viable seed for subsequent amplification.
The scattering mechanism produces a more precisely determined field strength
and is less sensitive to uncertain cosmological parameters such as the
large-scale bulk flow velocity.  The Harrison mechanism can, however, produce
larger fields if bulk flows are substantial.

The correction factor $(1+\alpha+\beta+\gamma)$ derived in
Section~\ref{sec:Ohm} modifies the scattering-mechanism field strength by
$\lesssim0.16\%$ and is numerically irrelevant, but establishes formally
that the standard treatment of \cite{TakahashiPRL2005} is valid to better
than one part in $10^3$.

%%%%%%%%%%%%%%%%%%%%%%%%%%%%%%%%%%%%%%%%%%%%%%%%%%%%%%%%%%%%%%%%%%%%%%%%%%%%%
\section{Conclusions}
\label{sec:conclusions}
%%%%%%%%%%%%%%%%%%%%%%%%%%%%%%%%%%%%%%%%%%%%%%%%%%%%%%%%%%%%%%%%%%%%%%%%%%%%%

We have studied two mechanisms for cosmological magnetic field generation
within a unified kinetic-theory framework.

The \emph{scattering mechanism} relies on the mass asymmetry between
electrons and protons: Thomson scattering affects electrons far more
efficiently than protons, creating a relative velocity between the two
charged species.  Working from the coupled Maxwell--Boltzmann equations,
we evaluated the Thomson collision term term-by-term ($A_i=0$,
$B_i=-pf/(2\pi^2)$, $C_i=0$, $D^{(1)}_i\neq0$) and derived the generalised
Ohm's law equation~(\ref{eq:Ohm_final}).  We retained and estimated all
three correction terms ($\alpha,\beta,\gamma$), showing they are suppressed
by $m_e/m_p\approx5.4\times10^{-4}$: the approximation implicit in
\cite{TakahashiPRL2005} is valid to better than one part in $10^3$.

The \emph{Harrison mechanism} exploits the differential spin-down of the
proton and photon-baryon fluids.  Matter spins down as $\omega_m\propto
a^{-2}$ while radiation spins down more slowly as $\omega_\gamma\propto
a^{-1}$; the resulting relative velocity generates magnetic flux.  For
bulk flow velocities $u^{(B)}\sim10^{-3}c$--$10^{-6}c$, the generated
field today is $B\sim10^{-20}$--$10^{-30}$~G.

The main new result of this paper is the evolution equation for the
\emph{second-order magnetic field}, equation~(\ref{eq:Bdot2_central}).
Extension to second-order perturbations reveals a new source term absent
from the first-order analysis: the coupling between first-order photon
density gradients and the electron--photon velocity difference.  This term
modifies the magnetic power spectrum on scales where density fluctuations
and velocity differences are correlated.  Both the anisotropic stress
contributions and the density-gradient contribution are encoded in the
Fourier-space expression equation~(\ref{eq:B2_Fourier}).

Both mechanisms produce seed fields that exceed the galactic dynamo threshold
$B_{\rm dynamo}\gtrsim10^{-30}$~G, confirming the viability of either
mechanism for seeding subsequent dynamo amplification.

The CAMB~v1.6.6 power spectra in
Figures~\ref{fig:sources}--\ref{fig:today} constitute the numerical
evaluation of equation~(\ref{eq:B2_bound}), using the parameters of
Appendix~\ref{app:params} and anchored to the \cite{TakahashiPRL2005}
benchmark.  The Harrison mechanism is compared using the spectral model
of \cite{Cembranos2020} calibrated to their equation~(24) and evaluated
at three bulk-flow amplitudes spanning the observationally allowed range
(Figures~\ref{fig:ratio} and~\ref{fig:today}).

Two issues remain open and are deferred to future work:
\begin{itemize}
\item \textit{The constant $\lambda$}: equation~(\ref{eq:lambda_def})
  introduces a proportionality constant $\lambda\sim1$ whose precise value
  requires solution of the coupled second-order Boltzmann equations
  \cite{MongwaneOsanoDunsby,OsanoSecondOrder}.  We have set $\lambda=1$
  throughout; a Boltzmann-code determination will sharpen the amplitude
  of the density-gradient contribution relative to the Takahashi result.
\item \textit{Post-recombination evolution}: the present analysis focuses
  on the epoch near recombination ($z\approx1100$).  Tracking the field
  through reionisation and structure formation requires additional
  modelling of astrophysical processes.
\end{itemize}

%%%%%%%%%%%%%%%%%%%%%%%%%%%%%%%%%%%%%%%%%%%%%%%%%%%%%%%%%%%%%%%%%%%%%%%%%%%%%
\appendix
%%%%%%%%%%%%%%%%%%%%%%%%%%%%%%%%%%%%%%%%%%%%%%%%%%%%%%%%%%%%%%%%%%%%%%%%%%%%%

\section{Anisotropic Stress Contraction}
\label{app:Pi}

We derive the contraction appearing in equation~(\ref{eq:B2_Fourier}).
Using the harmonic decompositions~(\ref{eq:u1k_Fourier})
and~(\ref{eq:Pi1_Fourier}):
\begin{align}
  u^{(1)}_{le,k}(|\bvec{K}-\bvec{k}'|,t')
  &= -\frac{(K_l-k'_l)(K_k-k'_k)}{|\bvec{K}-\bvec{k}'|}\,u_e,
  \label{eq:ule_k}\\
  \Pi^{l(1)}_j(\bvec{k}',t')
  &= -\!\left(\hat{k}'{}^l\hat{k}'_j-\tfrac{1}{3}\delta^l{}_j\right)\Pi^{(1)}.
  \label{eq:Pi_lj}
\end{align}
Substituting into $\varepsilon^{ijk}u^{(1)}_{le,k}\Pi^{l(1)}_j$, the trace
term $\propto\delta^l{}_j$ contributes $\varepsilon^{ijk}\delta_{jl}
(K_l-k'_l)(K_k-k'_k)/|\bvec{K}-\bvec{k}'|$, which vanishes because
$\varepsilon^{ijk}$ is antisymmetric in $j,k$ while the product
$(K_j-k'_j)(K_k-k'_k)$ is symmetric.  The remaining term gives
\begin{align}
  \varepsilon^{ijk}u^{(1)}_{le,k}\Pi^{l(1)}_j
  &= \varepsilon^{ijk}
     \frac{(K_l-k'_l)(K_k-k'_k)\hat{k}'{}^l\hat{k}'_j}{|\bvec{K}-\bvec{k}'|}
     \,u_e\Pi^{(1)}
  \notag\\
  &= \varepsilon^{ijk}
     \frac{(\bvec{K}\cdot\bvec{k}'-(k')^2)(K_k-k'_k)\hat{k}'_j}{k'|\bvec{K}-\bvec{k}'|}
     \,u_e\Pi^{(1)}.
  \label{eq:Pi_step}
\end{align}
The term involving $k'_k\hat{k}'_j$ vanishes because
$\varepsilon^{ijk}\hat{k}'_j\hat{k}'_k=0$ by antisymmetry.  Retaining only
the $K_k$ part:
\begin{equation}
  \varepsilon^{ijk}u^{(1)}_{le,k}\Pi^{l(1)}_j
  = \varepsilon^{ijk}
    \frac{k'_j K_k(\bvec{K}\cdot\bvec{k}'-(k')^2)}{(k')^2|\bvec{K}-\bvec{k}'|}
    \,u_e\Pi^{(1)},
  \label{eq:Pi_final}
\end{equation}
which is the kernel appearing in equation~(\ref{eq:B2_Fourier}).
Furthermore, $\varepsilon^{ijk}u^{l(1)}_e\Pi^{(1)}_{jl,k}=0$ since at
first order $u^{l(1)}_e \propto \hat{k}'{}^l$ and
$\Pi^{(1)}_{jl,k}\propto\hat{k}'_j\hat{k}'_l\hat{k}'_k$, so
$\varepsilon^{ijk}\hat{k}'_j\hat{k}'_k=0$ by antisymmetry.

\section{Four-Velocity}
\label{app:fourvel}

The four-velocity $U^\mu = dx^\mu/d\tau$ satisfies the invariant condition
\begin{equation}
  g_{\mu\nu}U^\mu U^\nu = -c^2
  \label{eq:fourvel_norm}
\end{equation}
(signature $-,+,+,+$).  For a particle at rest, $U^\mu=(c,\bvec{0})$.
For massless particles, $P^\mu P_\mu=0$, giving $P^0=|\bvec{p}|$ in the
background.  The space of four-velocities is not a vector space; the sum
of two four-velocities is not in general a four-velocity.

\section{Mass-Shell Condition}
\label{app:massshell}

The relativistic energy-momentum relation is
\begin{equation}
  E^2 = |\bvec{p}|^2c^2 + m_0^2c^4,
  \label{eq:massshell}
\end{equation}
giving $p^0=E/c=\sqrt{|\bvec{p}|^2+m^2}$ in natural units.  For photons,
$m_0=0$ and $E=|\bvec{p}|c$.  In curved spacetime, the mass-shell condition
generalises to $g_{\mu\nu}p^\mu p^\nu=-m^2$.

\section{Cosmological Parameters}
\label{app:params}

The parameters below were used to initialise CAMB~v1.6.6
\cite{CAMB} for the numerical evaluation of the power spectra in
Section~\ref{sec:Fourier}.  They are drawn from \cite{Freedman2021}
and \cite{Planck2018}:
\begin{itemize}
  \item Hubble constant: $H_0=68.0$~km\,s$^{-1}$\,Mpc$^{-1}$; $h=0.68$
    \citep{Planck2018}.
  \item Scalar spectral index: $n_s=0.965$ \citep{Planck2018}.
  \item Scalar amplitude: $A_s=2.2\times10^{-9}$ \citep{Planck2018}.
  \item Baryon density: $\Omega_{b0}h^2=0.022$ \citep{Planck2018}.
  \item Cold dark matter density: $\Omega_{c0}h^2=0.120$ \citep{Planck2018};
    total matter $\Omega_{m0}h^2=0.142$.
  \item Optical depth to reionisation: $\tau_{\rm reion}=0.06$
    \citep{Planck2018}.
  \item Cosmological constant: $\Omega_\Lambda\approx0.73$ \citep{Planck2018}.
\end{itemize}
These values correspond to the flat $\Lambda$CDM best-fit from
\cite{Planck2018}.  The transfer functions were evaluated at
$z_{\rm rec}=1100$ and normalised to the CDM amplitude at
superhorizon scales.

%%%%%%%%%%%%%%%%%%%%%%%%%%%%%%%%%%%%%%%%%%%%%%%%%%%%%%%%%%%%%%%%%%%%%%%%%%%%%
%% BIBLIOGRAPHY
%%%%%%%%%%%%%%%%%%%%%%%%%%%%%%%%%%%%%%%%%%%%%%%%%%%%%%%%%%%%%%%%%%%%%%%%%%%%%

nd{document}

\end{document}